\documentclass{aastex631}
\accepted{May 14, 2025}

\shorttitle{Molecular Complexity of G336 }
\shortauthors{Duan et al.}
\graphicspath{{./}{figures/}}

\begin{document}

\title{An ALMA Study of Molecular Complexity in the Hot Core G336.99-00.03 MM1}

\correspondingauthor{Qian Gou}
\email{qian.gou@cqu.edu.cn}

\author[0009-0002-5477-6309]{Chunguo Duan}
\affiliation{School of Chemistry and Chemical Engineering, Chongqing University, Daxuecheng South Rd. 55, Chongqing 401331, People’s Republic of China}
\affiliation{Chongqing Key Laboratory of Chemical Theory and Mechanism, Chongqing University, Daxuecheng South Rd. 55, Chongqing 401331, People’s Republic of China}

\author[0000-0003-3831-3582]{Qian Gou}
\affiliation{School of Chemistry and Chemical Engineering, Chongqing University, Daxuecheng South Rd. 55, Chongqing 401331, People’s Republic of China}
\affiliation{Chongqing Key Laboratory of Chemical Theory and Mechanism, Chongqing University, Daxuecheng South Rd. 55, Chongqing 401331, People’s Republic of China}

\author[0000-0002-5286-2564]{Tie Liu}
\affiliation{Shanghai Astronomical Observatory, Chinese Academy of Sciences, Nandan Rd. 80, Shanghai 200030, People’s Republic of China}
\affiliation{Key Laboratory for Research in Galaxies and Cosmology, Shanghai Astronomical Observatory, Chinese Academy of Sciences, Nandan Rd. 80, Shanghai 200030, People’s Republic of China}

\author[0000-0001-5950-1932]{Fengwei Xu}
\affiliation{Department of Astronomy, Peking University, 5 Yiheyuan Road, Haidian District, Beijing 100871, People’s Republic of China}
\affiliation{Kavli Institute for Astronomy and Astrophysics, Peking University, 5 Yiheyuan Road, Haidian District, Beijing 100871, People’s Republic of China}

\author[0000-0001-8514-6989]{Xuefang Xu}
\affiliation{School of Chemistry and Chemical Engineering, Chongqing University, Daxuecheng South Rd. 55, Chongqing 401331, People’s Republic of China}
\affiliation{Chongqing Key Laboratory of Chemical Theory and Mechanism, Chongqing University, Daxuecheng South Rd. 55, Chongqing 401331, People’s Republic of China}

\author{Junlin Lan}
\affiliation{School of Chemistry and Chemical Engineering, Chongqing University, Daxuecheng South Rd. 55, Chongqing 401331, People’s Republic of China}
\affiliation{Chongqing Key Laboratory of Chemical Theory and Mechanism, Chongqing University, Daxuecheng South Rd. 55, Chongqing 401331, People’s Republic of China}

\author[0000-0002-7237-3856]{Ke Wang}
\affiliation{Kavli Institute for Astronomy and Astrophysics, Peking University, 5 Yiheyuan Road, Haidian District, Beijing 100871, People’s Republic of China}

\author[0000-0002-3319-1021]{Laurent Pagani}
\affiliation{LUX, Observatoire de Paris, PSL Research University, CNRS, Sorbonne Universités, UPMC Univ. Paris 06, 75014 Paris, France}

\author[0000-0003-4811-2581]{Donghui Quan}
\affiliation{Research Center for Intelligent Computing Platforms, Zhejiang Laboratory, Hangzhou 311100, People’s Republic of China}

\author[0000-0001-6106-1171]{Junzhi Wang}
\affiliation{Guangxi Key Laboratory for Relativistic Astrophysics, Department of Physics, Guangxi University, Nanning 530004, People’s Republic of China}

\author[0000-0001-8315-4248]{Xunchuan Liu}
\affiliation{Shanghai Astronomical Observatory, Chinese Academy of Sciences, Nandan Rd. 80, Shanghai 200030, People’s Republic of China}
\affiliation{Key Laboratory for Research in Galaxies and Cosmology, Shanghai Astronomical Observatory, Chinese Academy of Sciences, Nandan Rd. 80, Shanghai 200030, People’s Republic of China}

\author{Mingwei He}
\affiliation{School of Chemistry and Chemical Engineering, Chongqing University, Daxuecheng South Rd. 55, Chongqing 401331, People’s Republic of China}
\affiliation{Chongqing Key Laboratory of Chemical Theory and Mechanism, Chongqing University, Daxuecheng South Rd. 55, Chongqing 401331, People’s Republic of China}



\begin{abstract}

High-mass star formation involves complex processes, with the hot core phase playing a crucial role in chemical enrichment and the formation of complex organic molecules. However, molecular inventories in hot cores remain limited. Using data from the ALMA Three-millimeter Observations of Massive Star-forming regions survey (ATOMS), the molecular composition and evolutionary stages of two distinct millimeter continuum sources in the high-mass star forming region G336.99-00.03 have been characterized. MM1, with 19 distinct molecular species detected, along with 8 isotopologues and several vibrationally/torsionally excited states, has been identified as a hot core. MM2 with only 5 species identified, was defined as a HII region. Isotopic ratios in MM1 were derived, with $^{12}$C/$^{13}$C ranging from 16.0 to 29.2, $^{16}$O/$^{18}$O at 47.7, and $^{32}$S/$^{34}$S at 19.2. Molecular abundances in MM1 show strong agreement with other sources and three-phase warm-up chemical models within an order of magnitude for most species. Formation pathways of key molecules were explored, revealing chemical links and reaction networks. This study provides a detailed molecular inventory of two millimeter continuum sources, shedding light on the chemical diversity and evolutionary processes in high-mass star-forming regions. The derived molecular parameters and isotopic ratios offer benchmarks for astrochemical models, paving the way for further investigation into the formation and evolution of complex organic molecules during the hot core phase.

\end{abstract}

 \keywords{Star formation (1569), Isotopic abundances (867), Complex organic molecules (2256), Interstellar medium (847)}


\section{Introduction} \label{sec:intro}

High-mass stars (M $\geq$ 8M$_{\odot}$) are born inside hot cores (HCs) embedded in massive molecular clouds \citep{Kur00}. These HCs are characterized by their compact sizes ($\leq$ 0.1 pc), elevated gas temperatures ($\geq$ 100 K), high gas densities (10$^{5}$-10$^{8}$ cm$^{-3}$), and numerous emission lines from complex organic molecules (COMs) \citep{Dis98, Hos09, Rat11}. Such environments are crucial for driving chemical complexity in the interstellar medium (ISM), attracting extensive scientific attention \citep{Shi21}. Investigating the molecular inventories and abundances in HCs is essential for elucidating the evolutionary pathways that drive complex chemistry.

Spectral surveys play a key role in uncovering the chemical composition of HCs. Surveys with broad bandwidth, which capture multiple transitions across a wide range of upper-level energies, provide valuable insights into the physical and chemical conditions shaping molecular emission. While many studies have conducted spectral surveys of HCs using single-dish telescopes \citep{Sch06, Ber10, Bel13, Col20, Liu22, Liu24b, Mar24} and millimeter/submillimeter interferometric arrays \citep{Gie21, Liu20, Liu24a}, only a limited number of sources — such as Sgr B2(N), OMC-1, G34.3+0.15, G9.62+0.19, and G328.25-0.53 — have been investigated in detail \citep{Sut95, Mac96, Bel16, Bel19, Bou22, Pen22}. Notably, the EMoCA (Exploring Molecular Complexity with ALMA) and ReMoCA (Re-exploring molecular complexity with ALMA) surveys of Sgr B2(N) \citep{Bel16, Bel19}, have significantly advanced our understanding of HC chemistry. Despite these advances, the formation and evolution of COMs in HCs remain challenging to decipher. Studying molecular inventories in HCs at varying evolutionary stages offers a promising pathway for addressing these challenges.

Herein, we report molecular inventories of two millimeter continuum sources in G336.99-00.03 that are at different stages of evolution. The high-mass star forming region G336.99-00.03 is located at a distance of 7.68 kpc from Earth and exhibits a bolometric luminosity of $\sim$2.5$\times$10$^{5}$ L$_{\odot}$ \citep{Urq18, Xu24}. Recently, surveys from “ALMA Three-millimeter Observations of Massive Star-forming regions” \citep[ATOMS,] []{Liu20} and “Querying Underlying mechanisms of massive star formation with ALMA-Resolved gas Kinematics and Structures” \citep[QUARKS,] []{Liu24a} at Band 3 and Band 6 have revealed that G336.99-00.03 hosts multiple massive clumps, including two millimeter continuum sources at distinct evolutionary stages. This makes G336.99-00.03 an ideal laboratory for exploring the formation and evolution of COMs. The brightest source, G336.99-00.03 MM1, exhibits rich chemical activity, including previously detected maser lines of OH and CH$_{3}$OH \citep{Sev97, Wal97, Wal98, Goe04} and emission lines of C$_{2}$H$_{5}$CN, CH$_{3}$OCHO, and CH$_{3}$OH \citep{Qin22}.

This study marks the first systematic survey of G336.99-00.03, providing an opportunity to examine the physical and chemical properties — such as temperature, density, and evolutionary stage — of its millimeter continuum sources. The layout of this paper is as follows: Sect. \ref{sec:sec2} outlines the observations and data reduction process, while Sect. \ref{sec:sec3} details the identification of molecular lines. In Sect. \ref{sec:sec4}, we present the results of continuum and spectral line surveys, deriving molecular parameters under the assumption of local thermodynamic equilibrium (LTE). Sect. \ref{sec:sec5} discusses the implications of our findings, focusing on the chemical links and formation pathways of the detected species. Finally, a summary of this study is given in Sect. \ref{sec:sec6}.

\section{Observation and data reduction} \label{sec:sec2}
\subsection{Observation} \label{sec:sec2.1}

\begin{figure}[ht!]
\centering
\includegraphics[width=0.8\columnwidth]{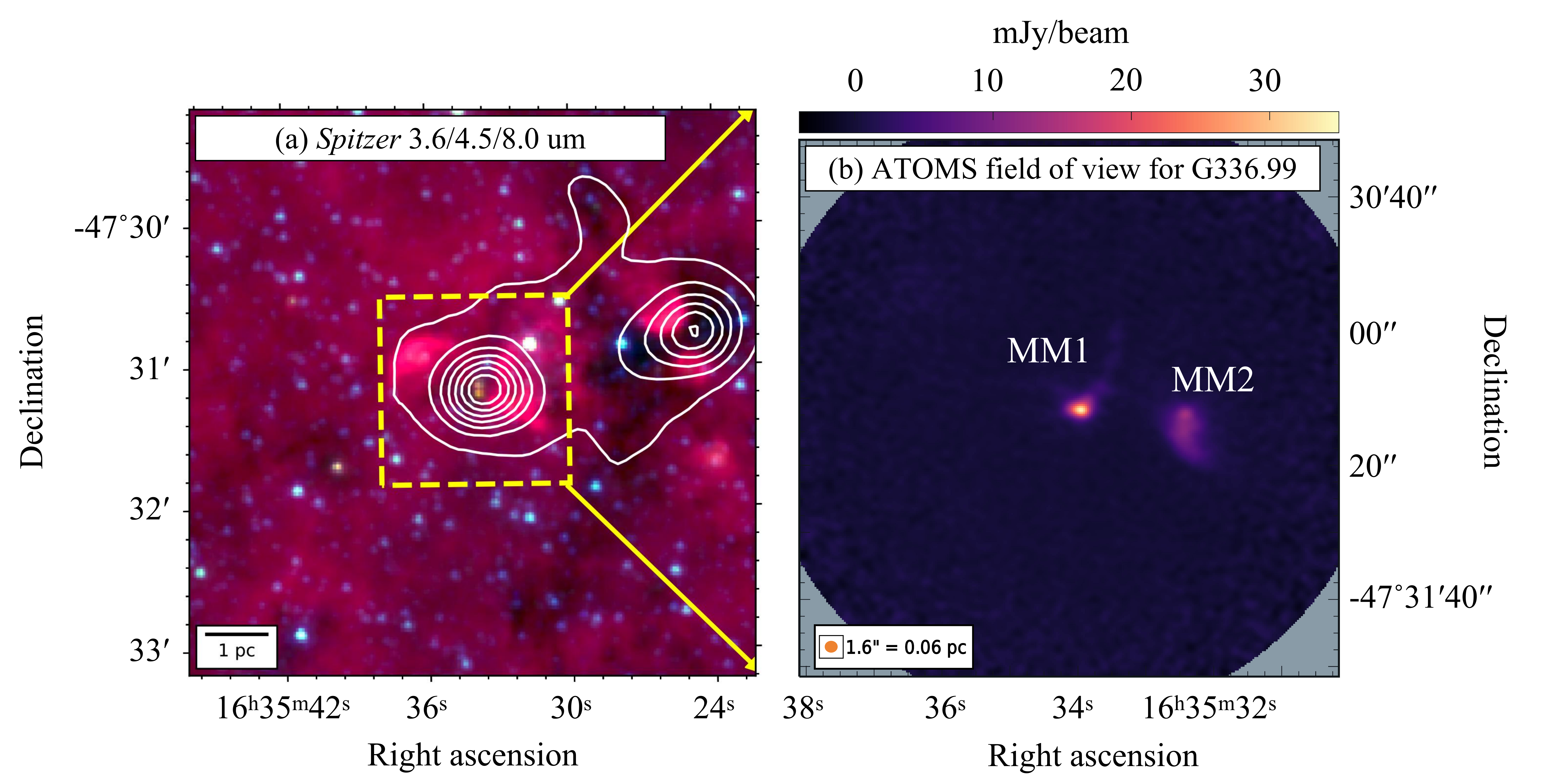}
\caption{(a) Spitzer pseudo-color map showing emissions at 3.6 $\mu$m (blue), 4.5 $\mu$m (green), and 8.0 $\mu$m (red). Overlaid white contours represent ATLASGAL 870 $\mu$m continuum emission at levels of 1.8, 3.0, 4.5, 5.9, 7.5, 9.0, and 10.7 Jy beam$^{-1}$. A 1 pc scale bar is shown in the lower left corner. (b) The ATOMS field of view of G336.99-00.03, with the 3 mm continuum emission as the background. The angular resolution, which is represented by the half-power beam width (HPBW), and the physical resolution scale are indicated in the lower left corner. The two identified sources, MM1 and MM2, are marked in white.
\label{fig:fig. 1}}
\end{figure}

G336.99-00.03 (IRAS 16318-4724) is one of the targets of the ATOMS (Project ID: 2019.1.00685.S; PI: Tie Liu), which targeted 146 bright infrared IRAS sources exhibiting bright CS (J = 2–1) emission \citep{Liu20}. Observations of this region were conducted using the Atacama Compact 7-m Array (ACA) at Band 3 on November 12, 2019, and the 12-m array on November 3, 2019. The on-source observation time is $\sim$5 min for the ACA and $\sim$3 min for the 12-m array, respectively. The phase center in both observations was set to RA(J2000) = $16^h 35^m 34^s$.11, Dec.(J2000) = $-47^\circ 31^\prime 11^{\prime \prime}$.3. Detailed observational parameters can be found in \citet{Liu20}. The survey utilized six spectral windows (SPWs) in the lower sideband, each with a bandwidth of 58.59 MHz with a spectral resolution of 0.2-0.4 km s$^{-1}$. These windows captured emissions from dense gas tracers such as the J = 1–0 transition of HCO$^{+}$ and the shock tracer SiO J = 2–1. Additionally, SPWs 7 and 8 offered a broad bandwidth of 1875 MHz with a spectral resolution of 1.6 km s$^{-1}$, primarily used in this work for continuum imaging and line survey. SPWs 7 and 8 covered frequencies ranging from 97.536 to 99.442 GHz and from 99.470 to 101.390 GHz, respectively, encompassing numerous COM emission lines.

Figure \ref{fig:fig. 1}(a) provides a global view of the region, represented by a Spitzer pseudo-color map combining emissions at 3.6 $\mu$m (blue), 4.5 $\mu$m (green), and 8.0 $\mu$m (red). The presence of infrared point sources and extended emission indicates active star formation. The ATLASGAL 870 $\mu$m emission contours, shown in white, reveal a dense gas reservoir with an estimated mass of approximately 5000 M$_{\odot}$ for the left clump \citep{Urq18}, confirming G336.99-00.03 as a pair of massive star-forming clumps. The second clump is clearly identified toward the west. It is dense and harbours bipolar outflows in infrared images.

Figure \ref{fig:fig. 1}(b) shows the ATOMS field of view of G336.99-00.03, highlighting the 3 mm continuum emission as the background. The region is resolved into two millimeter continuum sources, designated as MM1 and MM2, listed in order of decreasing right ascension (RA).

\subsection{Data reduction} \label{sec:sec2.2}
In this work, we utilized data from the 12-m array to identify COM lines within the HCs as their compact sizes in high mass star-forming regions and reduced susceptibility to missing flux make this approach suitable. Calibration of the visibilities data and imaging were performed using the CASA software package \citep[version 5.6,] []{Mcm07}. All images were corrected for primary beam response. Continuum images were generated from line-free channels in SPWs 7 and 8, centred at $\sim$99.4 GHz. Spectral line cubes for each SPW were created at their native spectral resolution. The synthesized beam size for the 3 mm continuum image of G336.99-00.03 is $1.70^{\prime \prime}$$\times$$1.49^{\prime \prime}$ with a position angle = $88.1^\circ$. The rms noise level is 0.3 mJy beam$^{-1}$ for the continuum and 4 mJy beam$^{-1}$ per channel for the spectral line data. Subsequently, the MAPPING application from the GILDAS\footnote{\url{http://www.iram.fr/IRAMFR/GILDAS}} software suite was employed to convert the data into Gildas format for detailed processing. Spectra were extracted for each core in a CLASS file format and subsequently converted from Jy beam$^{-1}$ to K for further analysis.

\section{Identification of molecular lines} \label{sec:sec3}

Under the assumption of the local thermodynamic equilibrium (LTE), LINEDB and WEEDS sets of routines inside CLASS \citep{Mar11} were used to identify and model the observed lines. Spectroscopic data from two major databases, the Cologne Database for Molecular Spectroscopy \citep[CDMS\footnote{\url{https://cdms.astro.uni-koeln.de}},] []{Mul01,Mul05} and the Jet Propulsion Laboratory database \citep[JPL\footnote{\url{https://spec.jpl.nasa.gov}},] []{Pic98}, provided the foundational molecular parameters required for line identification.

The spectral modeling relies on five key parameters: source size, line width, velocity offset, rotational temperature, and column density. In this work, the deconvolved angular sizes of the continuum sources were adopted as the source sizes. Line widths were derived through Gaussian fitting of the observed spectral lines. Velocity offsets were aligned using the CH$_{3}$OH line at 97582.798 MHz. For the two continuum sources, MM1 and MM2, the systemic velocities (V$_{\rm LSR}$) were measured as -119.8 km s$^{-1}$ and -117.3 km s$^{-1}$, respectively. Rotational temperatures and column densities were treated as variable parameters in the modeling process. These parameters were adjusted iteratively to ensure that the simulated spectra closely matched the observed lines, allowing for precise characterization of the molecular inventory.

\begin{figure}[h]
\centering
\includegraphics[width=0.8\columnwidth]{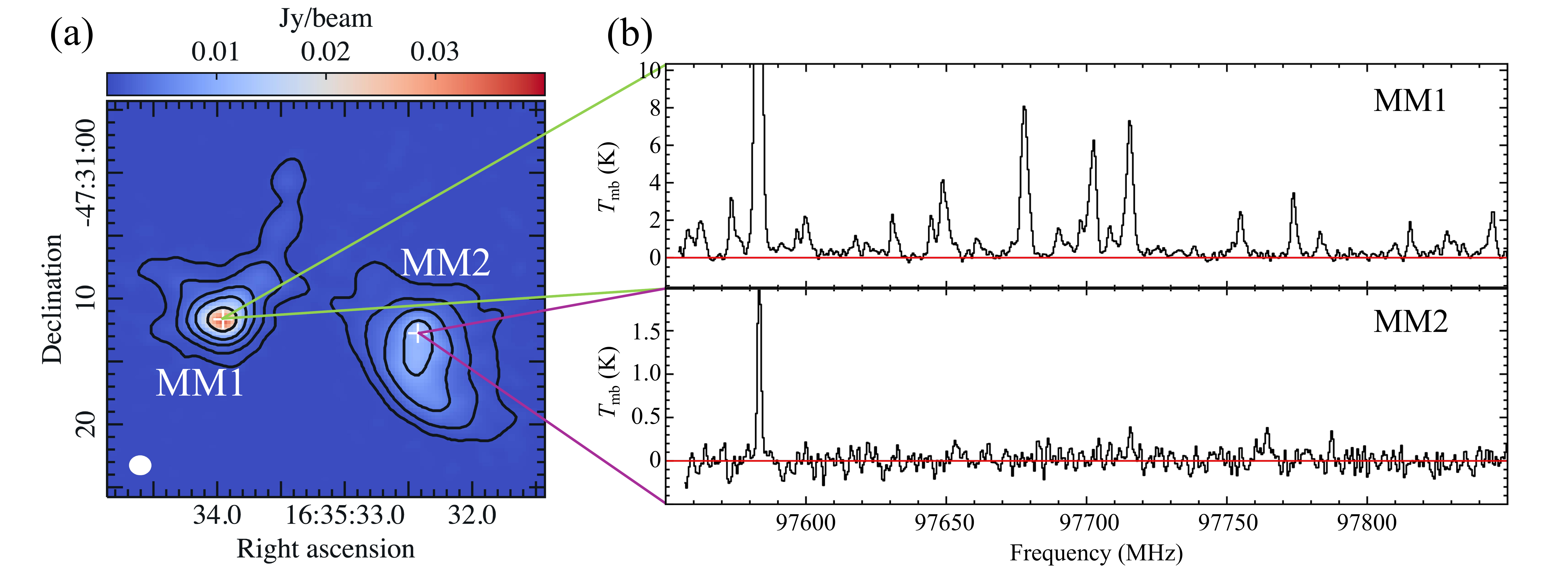}
\caption{(a) Three-millimeter continuum image of G336.99-00.03 obtained with ALMA band 3. Contour levels are drawn at (3, 6, 15, 30, 60)$\times \sigma$, where $\sigma$=0.3 mJy beam$^{-1}$ is the rms noise level of the 3 mm continuum image. The white ellipse indicates the synthesized beam ($1.70^{\prime\prime}$$\times$$1.49^{\prime\prime}$, $\mathrm{(position \ angle = 88.1^\circ}$)). (b) Observed spectra between 97.55 and 97.85 GHz towards MM1 (top panel) and MM2 (bottom panel). The brightness temperature scale (K) is given on the left.
\label{fig:fig. 2}}
\end{figure}

\begin{deluxetable*}{cccccc}
\tablenum{1}
\tablecaption{Parameters of continuum sources. \label{table:table 1}}
\tablewidth{0pt}
\tablehead{
\colhead{Name} & \colhead{R.A.} & \colhead{Dec} &
\colhead{Size} & \colhead{$I_{\rm peak}$} & \colhead{$S_{\rm \nu}$} \\
\colhead{} & \colhead{(hh:mm:ss)} & \colhead{(dd:mm:ss)} &
\colhead{($''$)} & \colhead{(mJy beam$^{-1}$)} & \colhead{(mJy)} 
}
\decimalcolnumbers
\startdata
MM1 & 16:35:33.96 & -47:31:11.7 & 2.31$\times$1.37 & 33.51$\pm$0.75 & 76.7$\pm$2.3 \\
MM2 & 16:35:32.42 & -47:31:12.8 & 3.12$\times$2.54 & 13.09$\pm$0.46 & 54.6$\pm$2.3 \\
\enddata
\tablecomments{The size of each core represents the deconvolved size, which was derived from 2D Gaussian fitting to the 3 mm continuum emission.}
\end{deluxetable*}

\section{Results} \label{sec:sec4}

\subsection{The detected molecular species} \label{sec:sec4.1}

The three-millimeter continuum emission of G336.99-00.03 (Figure \ref{fig:fig. 2}(a)) reveals two millimeter continuum sources: MM1 and MM2. Analysis of their line emission shows that MM1 exhibits intense and rich spectra, whereas MM2 is largely devoid of significant line emission (Figure \ref{fig:fig. 2}(b)). Table \ref{table:table 1} lists the parameters of these two sources, including their positions, deconvolved sizes, integrated flux density ($S_{\nu}$), and peak flux density ($I_{\rm peak}$), obtained through two-dimensional Gaussian fitting.

\begin{figure}[]
\plotone{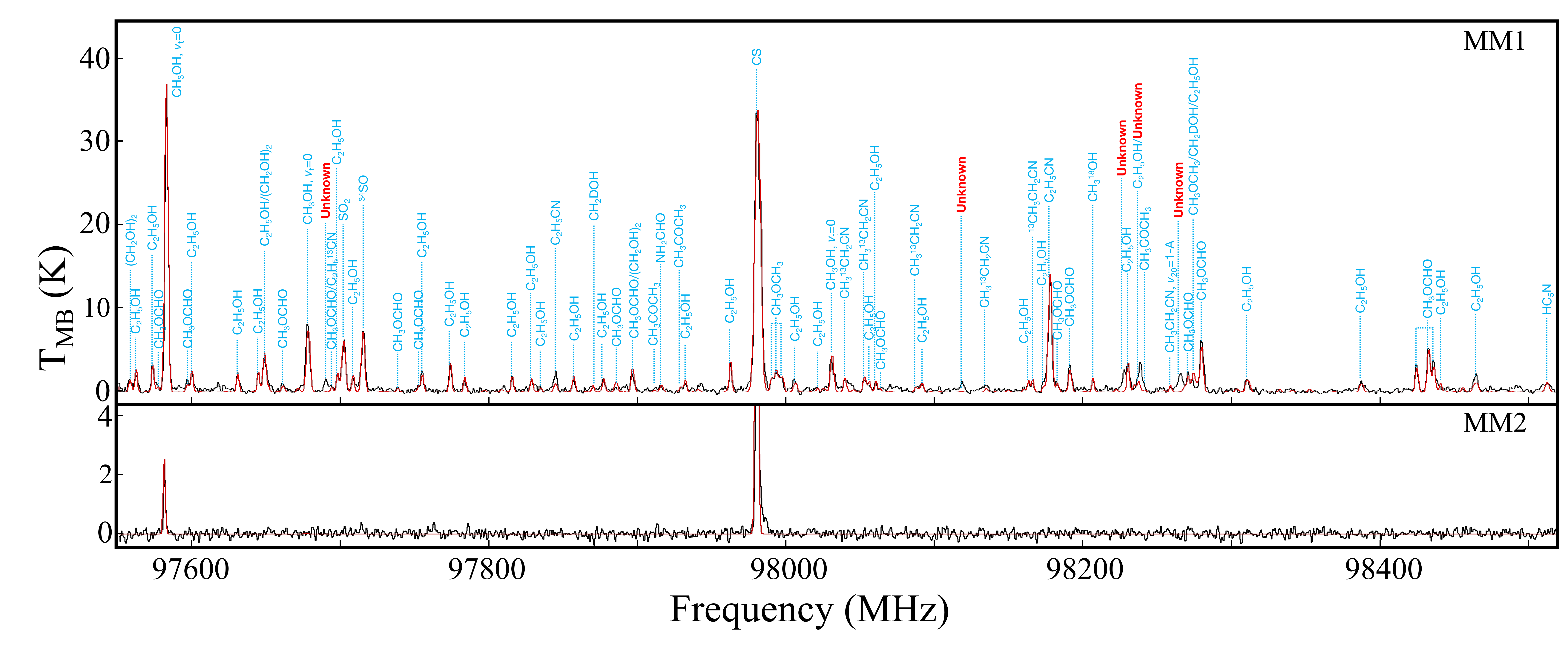}
\caption{Observed and synthetic spectra for MM1 and MM2. Black lines indicate the observed spectra, while red lines represent the modeled spectra. This segment covers the frequency range 97.55 - 98.52 GHz. Remaining frequencies are shown in Figure \ref{fig:fig. 12}. Unidentified lines are marked in red as “Unknown”.
\label{fig:fig. 3}}
\end{figure}

In total, 19 molecular species were detected in MM1 from over 300 transitions, while 5 species were identified in MM2 from only 7 transitions. Additionally, 8 isotopic species were detected in MM1. Notably, emission lines were observed from several vibrationally excited states of HC$_{3}$N ($\nu$$_{6}$=1, $\nu$$_{7}$=1, $\nu$$_{7}$=2, and $ \nu$$_{5}$=1/$\nu$$_{7}$=3) and C$_{2}$H$_{5}$CN ($\nu$$_{20}$=1-A), as well as torsionally excited CH$_{3}$OH ($\nu$$_{t}$=1). Table \ref{table:table 2} categorizes the detected species according to their number of atoms. The full-band beam-averaged spectra for MM1 and MM2 are displayed in Figures \ref{fig:fig. 3} and \ref{fig:fig. 12}, respectively, suggesting that the two sources may be at different evolution stages.

\begin{figure}[]
\centering
\includegraphics[width=0.6\columnwidth]{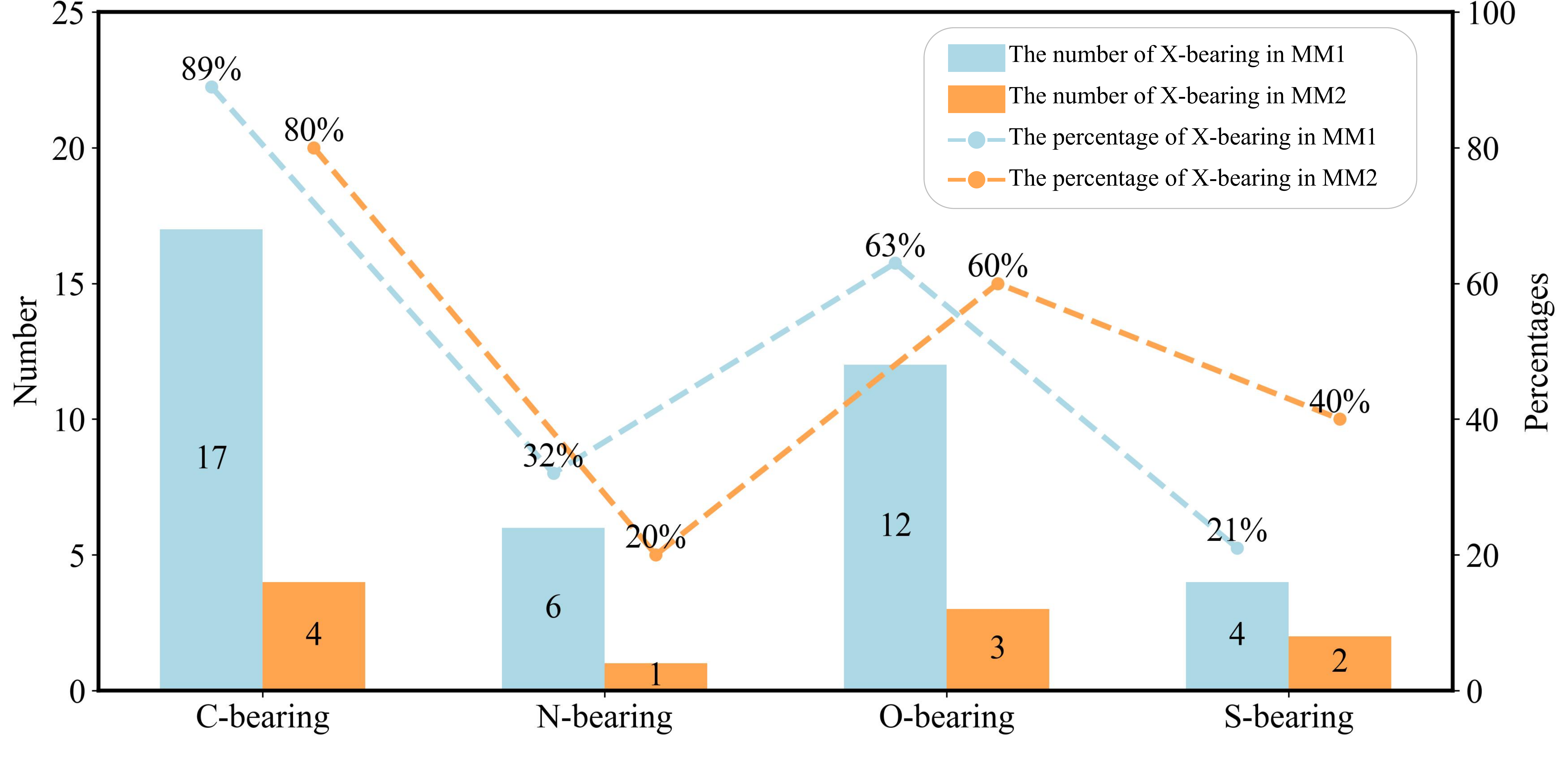}
\caption{Distribution of X-bearing species (X = C, N, O, S) in MM1 and MM2, shown as both number and percentage.
\label{fig:fig. 4}}
\end{figure}

Carbon (C)-bearing species were the most prevalent in MM1, accounting for 17 detected species, whereas sulfur (S)-bearing species were the least common, with only 4 identified. Nitrogen (N)-bearing and oxygen (O)-bearing species totaled 6 and 12, respectively, as shown in Figure \ref{fig:fig. 4}. In contrast, MM2 exhibited fewer line emissions, dominated by simple molecules. The transitions of CS J = 2–1, SO 2(3)–1(2), SO 5(4)–4(4), and HC$_{3}$N J = 11–10 were detected in MM2, along with two COMs, CH$_{3}$OH and CH$_{3}$CHO.

\begin{deluxetable*}{cccccccccc}
\tablenum{2}
\tablecaption{Detected molecular species in the MM1 and MM2. \label{table:table 2}}
\tablewidth{0pt}
\tablehead{
\colhead{} & \colhead{2 atoms} & \colhead{3 atoms} &
\colhead{4 atoms} & \colhead{5 atoms} & \colhead{6 atoms} & 
\colhead{7 atoms} & \colhead{8 atoms} & \colhead{9 atoms} & \colhead{10 atoms}
}
\decimalcolnumbers
\startdata
		{} & SO & SO$_{2}$ & NH$_{2}$D & H$_{2}$CCO & NH$_{2}$CHO & CH$_{3}$CHO & CH$_{3}$OCHO & C$_{2}$H$_{5}$CN &  CH$_{3}$COCH$_{3}$ \\
		{} & CS & {} & H$_{2}$CO & NH$_{2}$CN & CH$_{3}$NC & HC$_{5}$N & {} & CH$_{3}$$^{13}$CH$_{2}$CN &  (CH$_{2}$OH)$_{2}$ \\
		{} & $^{34}$SO & {} & {} & HC$_{3}$N & CH$_{3}$OH, $\nu$$_{t}$=0 & {} & {} & $^{13}$CH$_{3}$CH$_{2}$CN &  {}  \\
		{} & {} & {} & {} & HC$^{13}$CCN & CH$_{3}$OH, $\nu$$_{t}$=1 & {} & {} & C$_{2}$H$_{5}$CN, $\nu$$_{20}$=1-A &  {}  \\
		{} & {} & {} & {} & HCC$^{13}$CN & CH$_{3}$$^{18}$OH, $\nu$=0 & {} & {} & C$_{2}$H$_{5}$OH  &  {}  \\
		MM1 & {} & {} & {} & HC$_{3}$N,  $\nu$$_{6}$=1 & CH$_{2}$DOH & {} & {} & CH$_{3}$OCH$_{3}$  &  {}  \\
		{} & {} & {} & {} & HC$_{3}$N,  $\nu$$_{7}$=1 & CH$_{3}$SH, $\nu$=0 & {} & {} & {}  &  {}  \\
		{} & {} & {} & {} & HC$^{13}$CCN,  $\nu$$_{7}$=1 & CH$_{3}$SH, $\nu$=1 & {} & {} & {}  &  {}  \\
		{} & {} & {} & {} & HCC$^{13}$CN,  $\nu$$_{7}$=1 & {} & {} & {} & {}  &  {}  \\
		{} & {} & {} & {} & HC$_{3}$N,  $\nu$$_{7}$=2 & {} & {} & {} & {}  & {}  \\
		{} & {} & {} & {} & HC$_{3}$N, $ \nu$$_{5}$=1/$\nu$$_{7}$ =3 & {} & {} & {} & {}  &  {}  \\
		\hline	
		MM2 & SO & {} & {} & HC$_{3}$N & CH$_{3}$OH, $\nu$$_{t}$=0 & CH$_{3}$CHO & {} & {} &  {} \\	
		{} & CS & {} & {} & {} & {} & {} & {} & {} &  {} \\
		\hline	
\enddata
\end{deluxetable*}

\begin{deluxetable*}{cccccc}
\tablenum{3}
\tablecaption{Model fitting results of detected molecules in MM1. \label{table:table 3}}
\tablewidth{0pt}
\tablehead{
\colhead{Species} & \colhead{lines} &\colhead{$T_{\rm rot}$} & \colhead{$N_{\rm T}$} & \colhead{$\mathcal{\chi}_{{\rm H}_{2}}$} & \colhead{$\mathcal{\chi}_{{\rm CH}_{3} \rm OH}$} \\
\colhead{} & \colhead{} &\colhead{(K)} & \colhead{($\rm cm^{-2}$)} &
\colhead{} & \colhead{} 
}
\decimalcolnumbers
\startdata
SO                         &  2   & 156$\pm$1  &  $(2.05\pm$0.01)$\times 10^{17}$  &  $(1.15\pm$0.23)$\times 10^{-7}$    &   $(5.12\pm$0.03)$\times 10^{-2}$      \\ 
CS                         &  1   & [146]      &  $(5.39\pm$0.02)$\times 10^{16}$  &  $(3.01\pm$0.61)$\times 10^{-8}$    &   $(1.35\pm$0.01)$\times 10^{-2}$      \\ 
$^{34}$SO                  &  1   & [146]      &  $(1.07\pm$0.02)$\times 10^{16}$  &  $(5.98\pm$1.20)$\times 10^{-9}$    &   $(2.67\pm$0.01)$\times 10^{-3}$      \\ 
SO$_{2}$                   &  3   & 138$\pm$1  &  $(2.67\pm$0.01)$\times 10^{17}$  &  $(1.49\pm$0.30)$\times 10^{-7}$    &   $(6.68\pm$0.05)$\times 10^{-2}$      \\ 
NH$_{2}$D                  &  1   & [146]      &  $(7.34\pm$0.03)$\times 10^{15}$  &  $(4.10\pm$0.83)$\times 10^{-9}$    &   $(1.84\pm$0.01)$\times 10^{-3}$      \\ 
H$_{2}$CO                  &  2   & 249$\pm$2  &  $(3.40\pm$0.02)$\times 10^{17}$  &  $(1.90\pm$0.38)$\times 10^{-7}$    &   $(8.50\pm$0.06)$\times 10^{-2}$      \\ 
H$_{2}$CCO                 &  6   & 130$\pm$1  &  $(3.00\pm$0.03)$\times 10^{16}$  &  $(1.68\pm$0.34)$\times 10^{-8}$    &   $(7.50\pm$0.08)$\times 10^{-3}$      \\ 
NH$_{2}$CN                 &  6  & 135$\pm$24 &  $(2.31\pm$0.06)$\times 10^{15}$  &  $(1.29\pm$0.26)$\times 10^{-9}$    &   $(5.77\pm$0.15)$\times 10^{-4}$      \\ 
HC$_{3}$N                  &  1   & [146]      &  $(2.27\pm$0.02)$\times 10^{16}$  &  $(1.27\pm$0.26)$\times 10^{-8}$    &   $(5.68\pm$0.05)$\times 10^{-3}$      \\ 
HC$^{13}$CCN               &  1   & [146]      &  $(1.33\pm$0.01)$\times 10^{15}$  &  $(7.43\pm$1.49)$\times 10^{-10}$   &   $(3.33\pm$0.02)$\times 10^{-4}$      \\ 
HCC$^{13}$CN               &  1   & [146]      &  $(1.42\pm$0.01)$\times 10^{15}$  &  $(7.93\pm$1.60)$\times 10^{-10}$   &   $(3.55\pm$0.02)$\times 10^{-4}$      \\ 
NH$_{2}$CHO                &  4   & 191$\pm$2  &  $(1.94\pm$0.02)$\times 10^{17}$  &  $(1.08\pm$0.22)$\times 10^{-7}$    &   $(4.85\pm$0.06)$\times 10^{-2}$      \\ 
CH$_{3}$NC                 &  3   & 150$\pm$5  &  $(9.64\pm$0.32)$\times 10^{14}$  &  $(5.93\pm$1.10)$\times 10^{-10}$   &   $(2.41\pm$0.08)$\times 10^{-4}$      \\ 
CH$_{3}$OH, $\nu$$_{t}$ =0 &  4  & 162$\pm$1  &  $(4.00\pm$0.02)$\times 10^{18}$  &  $(2.23\pm$0.45)$\times 10^{-6}$    &               -                        \\ 
CH$_{3}$$^{18}$OH          &  2   & 149$\pm$1  &  $(8.39\pm$0.09)$\times 10^{16}$  &  $(4.69\pm$0.94)$\times 10^{-8}$    &   $(2.10\pm$0.03)$\times 10^{-2}$      \\ 
CH$_{2}$DOH                &  4  & 152$\pm$36 &  $(3.79\pm$0.91)$\times 10^{16}$  &  $(2.12\pm$0.67)$\times 10^{-8}$    &   $(9.47\pm$2.31)$\times 10^{-3}$      \\ 
CH$_{3}$SH, $\nu$ =0       &  3  & 100$\pm$8  &  $(1.43\pm$0.04)$\times 10^{16}$  &  $(7.99\pm$1.63)$\times 10^{-9}$    &   $(3.58\pm$0.11)$\times 10^{-3}$      \\ 
CH$_{3}$CHO                &  4  & 108$\pm$2  &  $(2.92\pm$0.08)$\times 10^{16}$  &  $(1.63\pm$0.33)$\times 10^{-8}$    &   $(7.30\pm$0.21)$\times 10^{-3}$      \\ 
HC$_{5}$N                  &  2   & 167$\pm$32 &  $(4.42\pm$0.36)$\times 10^{14}$  &  $(2.47\pm$0.54)$\times 10^{-10}$   &   $(1.11\pm$0.09)$\times 10^{-4}$      \\ 
CH$_{3}$OCHO               &  30  & 133$\pm$2  &  $(1.59\pm$0.04)$\times 10^{17}$  &  $(8.88\pm$1.80)$\times 10^{-8}$    &   $(3.98\pm$0.11)$\times 10^{-2}$      \\ 
C$_{2}$H$_{5}$CN           & 10  & 106$\pm$1  &  $(4.27\pm$0.03)$\times 10^{16}$  &  $(2.39\pm$0.48)$\times 10^{-8}$    &   $(1.07\pm$0.01)$\times 10^{-2}$      \\ 
CH$_{3}$$^{13}$CH$_{2}$CN  &  4  & 120$\pm$2  &  $(1.61\pm$0.22)$\times 10^{15}$  &  $(1.08\pm$0.41)$\times 10^{-9}$    &   $(4.85\pm$1.53)$\times 10^{-4}$      \\ 
$^{13}$CH$_{3}$CH$_{2}$CN  &  1   & [146]      & $(1.81\pm$0.02)$\times 10^{15}$  &  $(1.06\pm$0.21)$\times 10^{-9}$    &   $(4.75\pm$0.05)$\times 10^{-4}$      \\ 
C$_{2}$H$_{5}$OH           &  61  & 164$\pm$1  &  $(1.53\pm$0.01)$\times 10^{17}$  &  $(8.55\pm$1.72)$\times 10^{-8}$    &   $(3.82\pm$0.03)$\times 10^{-2}$      \\ 
CH$_{3}$OCH$_{3}$          &  12  & 121$\pm$4  &  $(6.68\pm$0.24)$\times 10^{17}$  &  $(3.73\pm$0.76)$\times 10^{-7}$    &   $(1.67\pm$0.06)$\times 10^{-1}$      \\ 
(CH$_{2}$OH)$_{2}$         &  13  & 213$\pm$9  &  $(1.06\pm$0.05)$\times 10^{17}$  &  $(5.92\pm$1.23)$\times 10^{-8}$    &   $(2.65\pm$0.13)$\times 10^{-2}$      \\ 
CH$_{3}$COCH$_{3}$         &  9  & 73$\pm$1   &  $(1.61\pm$0.04)$\times 10^{16}$  &  $(8.99\pm$1.82)$\times 10^{-9}$    &   $(4.02\pm$0.10)$\times 10^{-3}$      \\ 
\enddata
\tablecomments{Column (2): The lines represent the number of clean or slightly blended transitions detected for each molecule.}
\end{deluxetable*}

\subsection{Excitation temperatures, column densities, and abundances relative to H$_{2}$} \label{sec:sec4.2}

This study pays particular attention towards MM1, the source with the brightest continuum emission and chemical complexity in G336.99-00.03. Molecules with more than three detected transitions across a broad range of upper energy levels (E$_{u}$) were selected to derive excitation temperatures and total column densities. The derived values are summarized in Table \ref{table:table 3}. For molecules with only one or two detected transitions, the excitation temperatures were fixed at 146 K, assuming consistency with the average gas temperature derived for MM1, to facilitate the calculation of their column densities.

The column density of molecular hydrogen ($\mathit{N}_{{\rm H}_{2}}$) was estimated under the assumption of optically thin emission from dust, using the following equation \citep[e.g.] [] {Mar11, Bon19}:
\begin{equation}
N_{{\rm H}_{2}} = \frac{{S_{\nu}}{R_{gd}}}{{\mu}{m_{H}}{\Omega}{{\kappa}_{\nu}}{B_{\nu}}({T_{dust}})}
\label{eq:equation1}
\end{equation}
where $S_{\rm \nu}$ represents the total integrated continuum flux, ${R_{\rm gd}}$ is the gas-to-dust mass ratio (assumed to be 100), and ${\mu}$ is the mean molecular mass per {\rm H$_{2}$} molecule, taken as 2.8 \citep{Kau08}. ${m_{\rm H}}$ is the mass of a hydrogen atom, ${\Omega}$ is the beam solid angle. ${{\kappa}_{\nu}}$ is the dust opacity, assumed to be 0.2 cm$^{2}$g$^{-1}$ at ${\rm \lambda}$${\sim}$3 mm, following the OH5 dust model \citep{Oss94}. ${{B_{\nu}}({T_{\rm dust}})}$ is the Planck function at the dust temperature $T_{\rm dust}$, assumed to be equal to the average derived rotational temperature. The estimated column density of {\rm H$_{2}$} for the MM1 source is (1.79$\pm$0.36)$\times$10$^{24}$ cm$^{2}$, assuming an uncertainty of 20\%. The molecular abundance relative to {\rm H$_{2}$} ($\chi$ = $N_{\rm T}$/$\mathit{N}_{{\rm H}_{2}}$) was then estimated using the derived total column density of each molecule ($N_{\rm T}$) and the column density of molecular hydrogen ($\mathit{N}_{{\rm H}_{2}}$). The results are provided in Table \ref{table:table 3}.

\subsection{Isotopic ratio} \label{sec:sec4.3}

This section focuses on the isotopic ratios of carbon ($^{12}$C/$^{13}$C), oxygen ($^{16}$O/$^{18}$O), and sulfur ($^{32}$S/$^{34}$S), as summarized in Table \ref{table:table 4}. These ratios provide critical insights into the chemical processes occurring within HCs and the cycling of materials in ISM. The ratio $^{12}$C/$^{13}$C, computed from five species, is found to range from approximately 16.0 to 29.2. In contrast, the ratios of $^{16}$O/$^{18}$O and $^{32}$S/$^{34}$S were each derived from a single species, with values of ${\sim}$47.7 and ${\sim}$19.2, respectively. The $^{12}$C/$^{13}$C ratios derived from $^{13}$CH$_{3}$CH$_{2}$CN and the $^{13}$C isotopologues of HC$_{3}$N should be interpreted with caution, as only one clean transition was detected for each of these species. Future observations targeting multiple transitions would improve the reliability of these estimates.

It is worth noting that reliable isotope ratios can only be derived using lines with sufficiently low optical depths ($\tau$ $\ll$ 1). We have indeed noted that the anomalous $^{12}$C/$^{13}$C and $^{16}$O/$^{18}$O ratios derived from HC$_{3}$N and CH$_{3}$OH could be influenced by the high optical depths of the main isotopolog transitions. For each line used to calculate isotope ratios, the optical depth ($\tau$) computed with CLASS is listed in Table \ref{table:table B1}. The main isotopolog transitions for HC$_{3}$N and CH$_{3}$OH exhibit the high optical depths. We evaluated how source size assumptions affect opacity estimates, as the line-emitting regions ($\sim$ $1.69^{\prime \prime}$ for HC$_{3}$N and $\sim$ $1.61^{\prime \prime}$ for CH$_{3}$OH) are slightly more compact than the continuum deconvolved size used in our analysis. Using a larger source size could underestimate the opacities of the main isotopologues, which would lead to a slight underestimation of the $^{12}$C/$^{13}$C and $^{16}$O/$^{18}$O ratios. However, this effect is small and does not fully explain the observed low ratios. Therefore, we have retained the use of the continuum deconvolved size for consistency across all species, but note that the actual ratios may be slightly higher than reported.

The significance of these isotopic ratios and their implications for chemical evolution and isotopic fractionation within the studied regions are extensively discussed in Section \ref{sec:sec5.1}.

\subsection{Spatial distribution of the observed molecules} \label{sec:sec4.4}
\begin{figure}[]
\centering
\includegraphics[width=0.7\columnwidth]{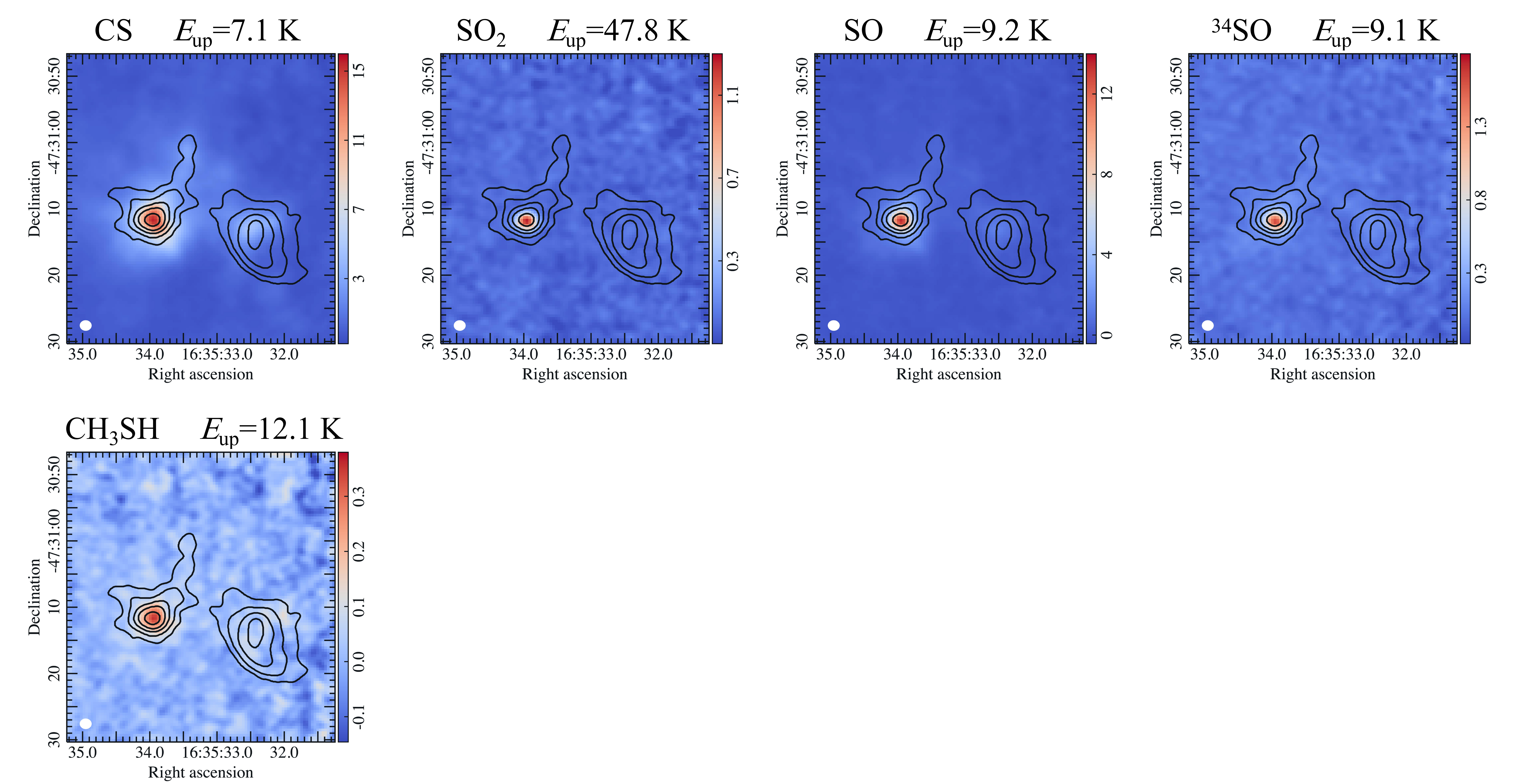}
\caption{Integrated emission maps of S-bearing molecules in G336.99-00.03, overlaid with 3 mm continuum emission contours. Contour levels are drawn at (3, 6, 15, 30, 60)$\times \sigma$, where $\sigma$=0.3 mJy beam$^{-1}$ is the rms noise level of the 3 mm continuum image. The solid ellipse in the bottom left corner indicates the synthesized beam for the continuum. Each panel displays the molecule name and upper energy level energy E$_{u}$ above the map. The unit of the color bar on the right is Jy beam$^{-1}$.
\label{fig:fig. 5}}
\end{figure}

\begin{figure}[]
\centering
\includegraphics[width=0.7\columnwidth]{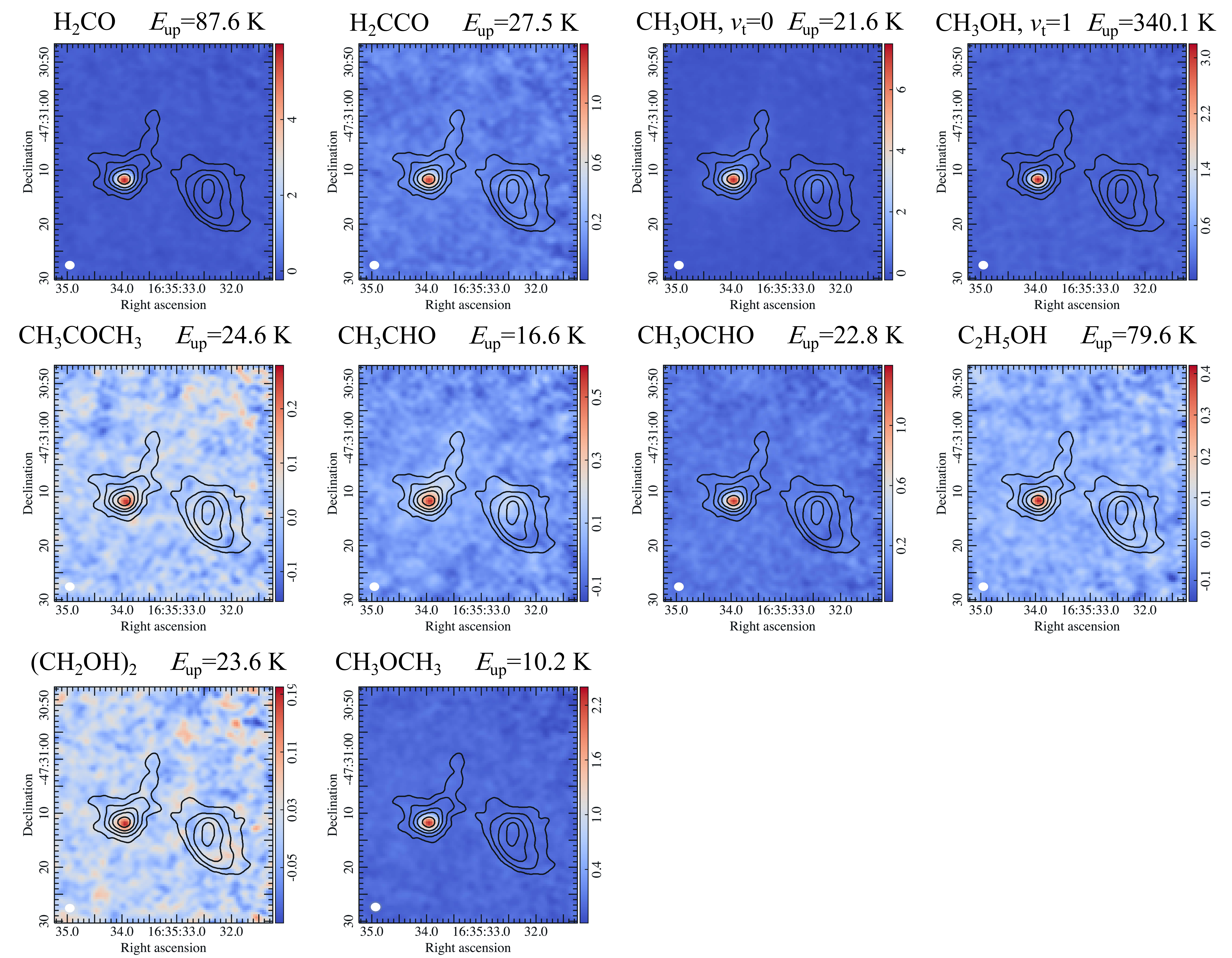}
\caption{Integrated emission maps of O-bearing molecules in G336.99-00.03, overlaid with 3 mm continuum emission contours. Details are as described in Figure \ref{fig:fig. 5}.
\label{fig:fig. 6}}
\end{figure}

\begin{figure}[]
\centering
\includegraphics[width=0.7\columnwidth]{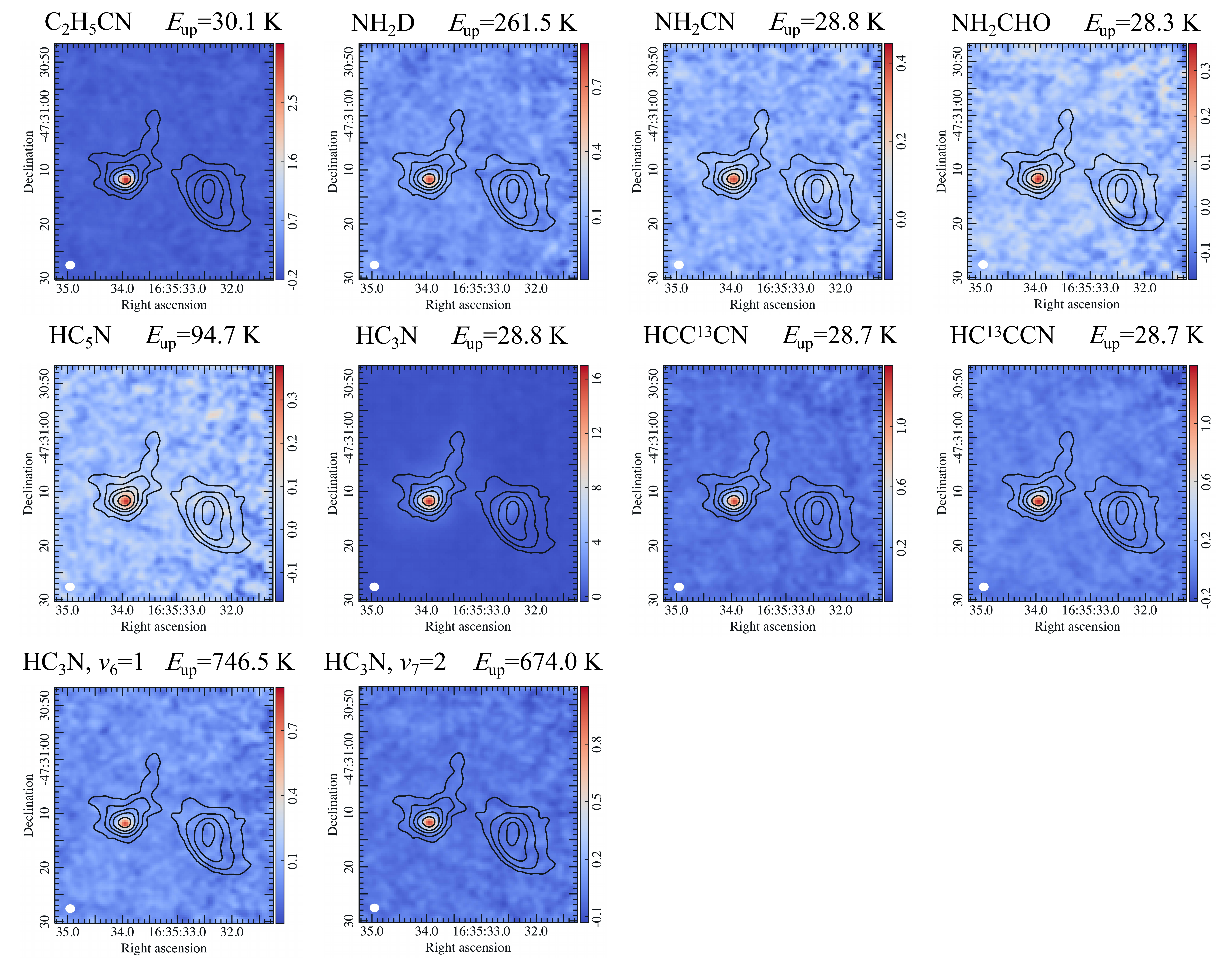}
\caption{Integrated emission maps of N-bearing molecules in G336.99-00.03, overlaid with 3 mm continuum emission contours. Details are as described in Figure \ref{fig:fig. 5}
\label{fig:fig. 7}}
\end{figure}

The spatial distributions of nitrogen (N)-, oxygen (O)-, and sulfur (S)-bearing molecules were analyzed using the Cube Analysis and Rendering Tool for Astronomy \citep[CARTA,] []{Com21}. Integrated intensity (moment 0) maps were generated for these molecules to investigate their spatial patterns and potential chemical pathways. Figures \ref{fig:fig. 5}, \ref{fig:fig. 6}, and \ref{fig:fig. 7}, show the moment 0 maps of S-, O-, and N-bearing molecules, respectively, overlaid on the 3 mm continuum emission. To ensure accuracy, all transitions utilized for generating the moment 0 maps were carefully verified to avoid contaminations from blending with other spectral lines. Molecules with weak emissions, such as CH$_{2}$DOH, were excluded from the moment 0 maps due to insufficient signal strength for reliable spatial representation. It can be noted that the emission of molecules in G336.99-00.06 appears highly compact, with most species showing nearly identical spatial distributions. This compactness, combined with the limited angular resolution of our observations, makes it challenging to detect subtle spatial variations. Future observations with higher angular resolution will be essential to probe potential spatial differences and provide deeper insights into the chemical structure of this source.

\begin{deluxetable*}{cc}
\tablenum{4}
\tablecaption{Isotopic ratios of carbon ($^{12}$C/$^{13}$C), oxygen ($^{16}$O/$^{18}$O), and sulfur ($^{32}$S/$^{34}$S) in MM1. \label{table:table 4}}
\tablewidth{0pt}
\tablehead{
\colhead{Species} & \colhead{Ratio} 
}
\decimalcolnumbers
\startdata
HC$_{3}$N/HC$^{13}$CCN & 17.1 $\pm$ 0.1 \\
HC$_{3}$N/HCC$^{13}$CN & 16.0 $\pm$ 0.1 \\
HC$_{3}$N $\nu$$_{7}$=1/HCC$^{13}$CN $\nu$$_{7}$=1 & 29.2 $\pm$ 1.2 \\
C$_{2}$H$_{5}$CN/CH$_{3}$$^{13}$CH$_{2}$CN & 26.5 $\pm$ 3.6 \\
C$_{2}$H$_{5}$CN/$^{13}$CH$_{3}$CH$_{2}$CN & 23.6 $\pm$ 0.3 \\
CH$_{3}$OH/CH$_{3}$$^{18}$OH & 47.7 $\pm$ 0.6 \\		
SO/$^{34}$SO & 19.2 $\pm$ 0.1 \\
\enddata
\end{deluxetable*}

\section{Discussion} \label{sec:sec5}

\subsection{Isotopic ratio} \label{sec:sec5.1}
Isotopic abundance ratios are critical for understanding the chemical evolution of the Milky Way \citep{Wil94}. Among these, the $^{12}$C/$^{13}$C ratio has been extensively studied in interstellar environments. In the central molecular zone (CMZ), this ratio typically ranges from 20 to 25 \citep[e.g.] []{Mil05, Riq10, Bel13, Hal17, Hum20}. Regions closer to the Galactic center generally exhibit lower ratios, while higher ratios ($\sim$50) are observed further from the center, as derived from H$_{2}$$^{12}$CO/H$_{2}$$^{13}$CO measurements \citep{Hen85}. Numerous studies have demonstrated that the $^{12}$C/$^{13}$C ratio varies systematically with the Galactocentric distance $\mathit{D}_{\rm GC}$ \citep{Lan90, Lan93}. A recent linear fit to the $^{12}$C/$^{13}$C ratio across 112 sources, based on H$_{2}$$^{12}$CO/H$_{2}$$^{13}$CO observations, was proposed by \citet{Yan19}:
\begin{equation}
^{12}\mathrm{C}/^{13}\mathrm{C} = (5.08 \pm 1.10) \mathit{D}_{\rm GC} (\rm kpc) + (11.86 \pm 6.60)
\label{eq:equation2}
\end{equation}

\begin{figure}[h]
\centering
\includegraphics[width=0.5\columnwidth]{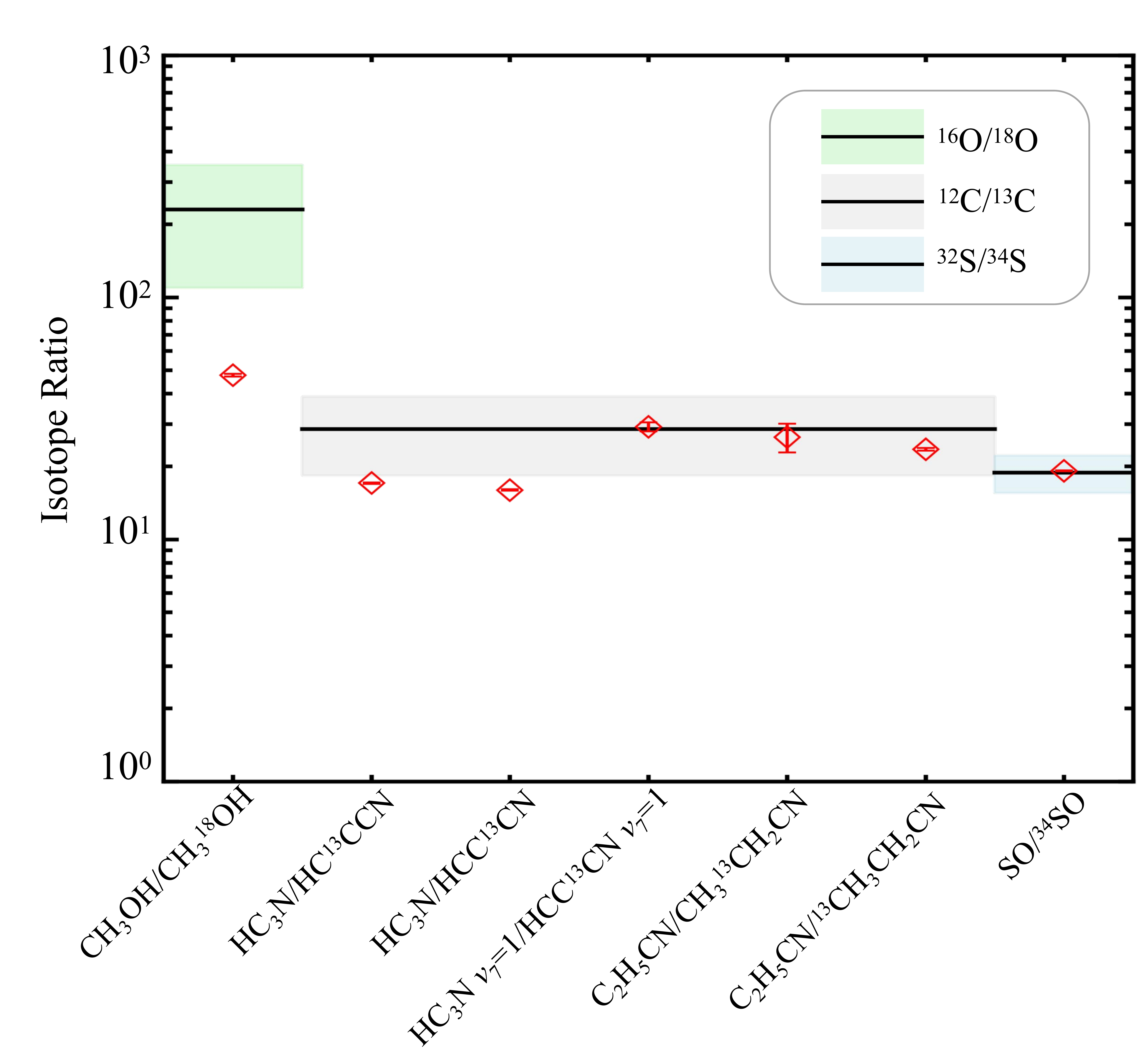}
\caption{Isotopic ratios of $^{12}$C/$^{13}$C, $^{32}$S/$^{34}$S and $^{16}$O/$^{18}$O at $\mathit{D}_{\rm GC}$=3.3 kpc, derived using Equations by \citet{Yan19, Yan23, Wil99}. Solid lines indicate first-order polynomial fits, with gray-, blue-, green-shaded areas showing the 1$\sigma$ intervals. Hollow diamonds represent isotopic ratios determined in this work, with the red error bars showing 1$\sigma$ uncertainties.
\label{fig:fig. 8}}
\end{figure}

At the Galactocentric distance of G336.99-00.03 \citep[$\mathit{D}_{\rm GC}$=3.3 kpc,] []{Liu20}, this equation predicts a $^{12}$C/$^{13}$C ratio of 28.6$\pm$10.2. Using HC$_{3}$N and C$_{2}$H$_{5}$CN isotopologues, the $^{12}$C/$^{13}$C ratio was estimated as reported in Table \ref{table:table 4}. These values, along with predictions from Equation by \citet{Yan19}, are plotted in Figure \ref{fig:fig. 8}. For C$_{2}$H$_{5}$CN, the derived ratios of 23.6 and 26.5 align well with Equation \ref{eq:equation2}. However, for HC$_{3}$N ($\nu$=0), the derived $^{12}$C/$^{13}$C ratios are 16.0 and 17.1, respectively, about 40\% lower than predicted. This discrepancy may arise from several factors. First, the large optical depth of HC$_{3}$N lines could lead to an underestimation of the $^{12}$C/$^{13}$C ratio, as optically thick lines do not accurately trace the true column density of the main isotopologue. Second, \citet{Yan19} highlighted that uncertainties such as distance effects, beam sizes variations, excitation temperature, isotope selective photodissociation, and chemical fractionation could influence the derived $^{12}$C/$^{13}$C ratios. While they argue that distance effects and excitation temperature variations are likely negligible, isotope selective photodissociation and chemical fractionation could play a role, particularly in environments with strong UV radiation or complex chemical pathways. For example, isotope selective photodissociation tends to increase the $^{12}$C/$^{13}$C ratio in photon-dominated regions (PDRs), whereas chemical fractionation could either enhance or deplete $^{13}$C depending on the dominant formation mechanism. In the case of HC$_{3}$N, the lower $^{12}$C/$^{13}$C ratio observed in our study may suggest that local environmental factors, such as chemical fractionation or variations in the molecular formation/destruction pathways, are influencing the isotopic ratios in G336.99-00.03 MM1. The vibrational excited states of HC$_{3}$N and its $^{13}$C isotopologues could provide a more robust $^{12}$C/$^{13}$C ratio, as these lines are less affected by optical depth effects. The $^{12}$C/$^{13}$C ratio derived from HC$_{3}$N $\nu$$_{7}$=1 and HCC$^{13}$CN $\nu$$_{7}$=1 is 29.2, which is in good agreement with the value derived from Equation \ref{eq:equation2}.

A similar linear relationship for $^{32}$S/$^{34}$S was proposed by \citet{Yan23}, based on $^{13}$C$^{32}$S/$^{13}$C$^{34}$S measurements from 110 high-mass star-forming regions:
\begin{equation}
^{32}\mathrm{S}/^{34}\mathrm{S} = (0.73 \pm 0.36) \mathit{D}_{\rm GC} (\rm kpc) + (16.50 \pm 2.07)
\label{eq:equation3}
\end{equation}
For oxygen, a fit based on H$_{2}$$^{12}$C$^{18}$O/H$_{2}$$^{13}$C$^{16}$O measurements was proposed by \citet{Wil99}.
\begin{equation}
^{16}\mathrm{O}/^{18}\mathrm{O} = (58.80 \pm 11.80) \mathit{D}_{\rm GC} (\rm kpc) + (37.10 \pm 82.60)
\label{eq:equation4}
\end{equation}

For G336.99-00.03, the calculated $^{32}$S/$^{34}$S and $^{16}$O/$^{18}$O ratios are 18.9$\pm$3.3 and 231.1$\pm$121.5, respectively, as shown in Figure \ref{fig:fig. 8}. The $^{32}$S/$^{34}$S ratio agrees well with Equation \ref{eq:equation3}. However, the derived $^{16}$O/$^{18}$O ratio is about 4 times lower than the value predicted from Equation \ref{eq:equation4}. This discrepancy may indicate anomalous isotopic fractionation in this region or a high CH$_{3}$OH opacity, potentially caused by a cold front layer. The absence of $^{13}$CH$_{3}$OH detection prevents a definitive conclusion. Finally, the small sample size introduces significant uncertainty in this analysis, emphasizing the need for additional observations to confirm these findings.

\begin{figure}[]
\plotone{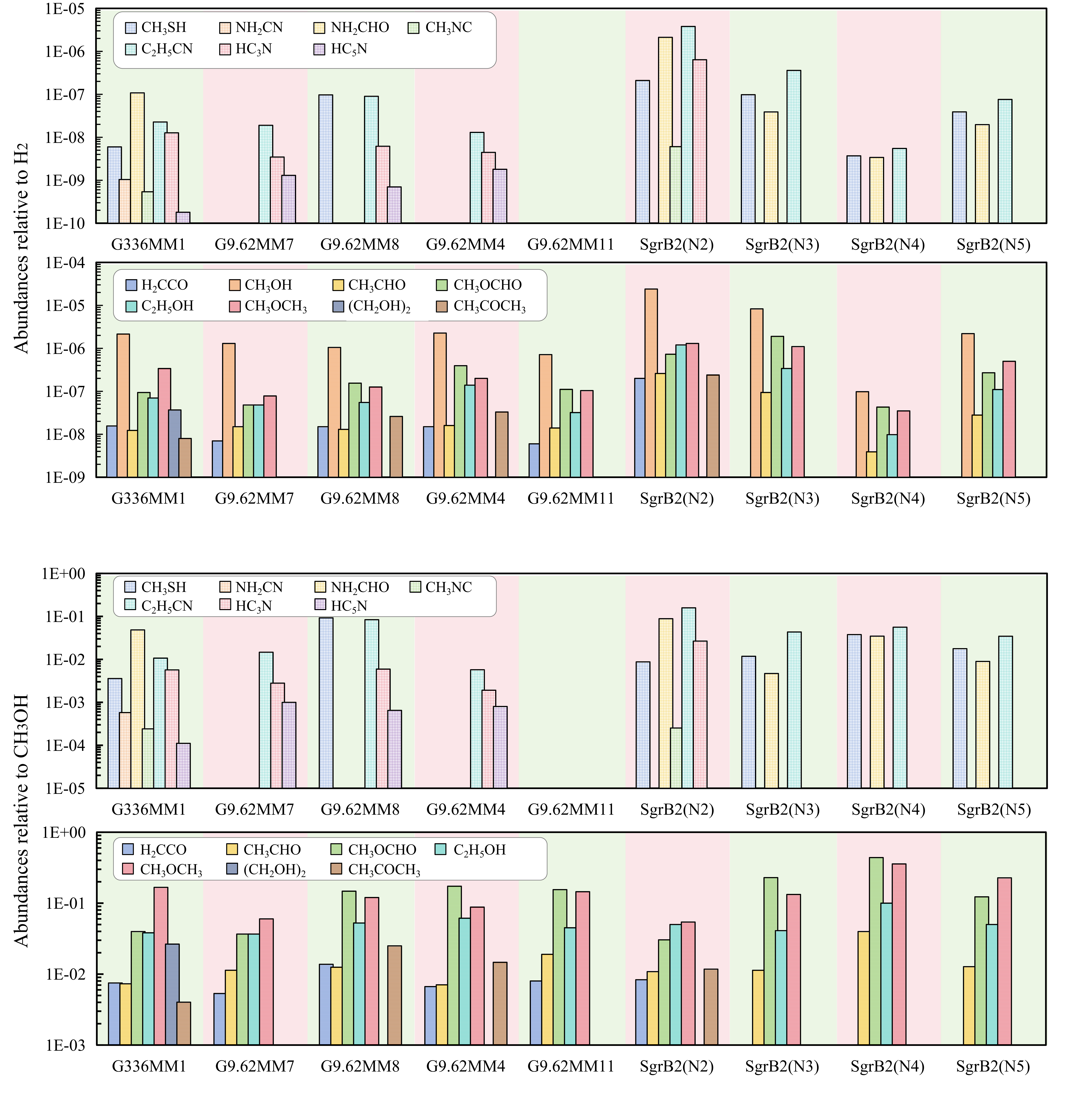}
\caption{Upper panel: Abundances of S-bearing, N-bearing and O-bearing molecules relative to H$_{2}$ in G336.999-00.03 MM1 and other sources. Lower panel: Abundances of S-bearing, N-bearing and O-bearing molecules relative to CH$_{3}$OH in G336.999-00.03 MM1 and other sources. All abundances are listed in Tables \ref{table:table C1} and \ref{table:table C2}. The source labeled as G336MM1 represents G336.99-00.03 MM1, while G9.62MM4, MM7, MM8, and MM11 represent the four hot cores in G9.62+0.19.
\label{fig:fig. 9}}
\end{figure}

\subsection{Evolution of two millimetre continuum sources} \label{sec:sec5.2}
Interstellar molecules, especially COMs, are valuable probes for studying the ISM. Their prevalence under specific physical conditions makes them essential tools for exploring the environments and evolutionary stages of different sources. Molecules in vibrationally or torsionally excited states, with higher upper energy levels (E$_{u}$), are typically reliable indicators of regions with higher kinetic temperatures and more advanced evolutionary stages \citep{Gie21}. For instance, the CH$_{3}$OH $\nu$$_{t}$=1 (6$_{1,6}$-5$_{0,5}$) line, with an upper-level energy of 340.1 K, indicates that the dense core has undergone substantial heating and is likely in the hot core or ultracompact HII (UCHII) phase \citep{Liu17}. Based on the detection of the CH$_{3}$OH $\nu$$_{t}$ =1 line, the evolutionary stages of the two millimeter continuum sources in G336.99-00.03 are classified as follows: The detection of numerous molecules, including those in vibrationally and torsionally excited states (e.g. HC$_{3}$N  $\nu$$_{7}$=1, HC$_{3}$N $\nu$$_{6}$=1, CH$_{3}$OH $\nu$$_{t}$=1), suggests that MM1 is in the HC phase. The detection of H$_{40 \alpha}$ and H$_{50 \beta}$ recombination lines indicates that MM2 has evolved into the HII stage. In this phase, ultraviolet radiation destroys many molecules, significantly reducing their detectability \citep{Pen22}.

\subsection{Chemical complexity toward G336.99-00.03 MM1} \label{sec:sec5.3}
In the previous section, we identified G336.99-00.03 MM1 as being in hot core stage. To further explore the chemical complexity in the hot core G336.99-00.03 MM1, it is informative to compare its molecular abundances with those of other sources and chemical models. In this section, we compare the molecular abundances of MM1 with data from key studies, including: 1) the ATOMS project of ALMA towards four hot cores at different evolutionary stages in G9.62+0.19 \citep{Pen22}; 2) ALMA observations of Galactic center sources Sgr B2(N2) \citep{Bel16}, Sgr B2(N3), Sgr B2(N4), and Sgr B2(N5) \citep{Bon17, Bon19}. The molecular abundances relative to H$_{2}$ and CH$_{3}$OH for each source, as well as those derived from chemical models are provided in Tables \ref{table:table C1} and \ref{table:table C2}.

\subsubsection{Comparison with other sources} \label{sec:sec5.3.1}
The upper panel of Figure \ref{fig:fig. 9} shows the molecular abundances relative to H$_{2}$ for the sources listed in Table \ref{table:table C1}. The abundances of N-bearing molecules are generally an order of magnitude lower than those of O-bearing molecules, consistent with the findings of \citet{Gel20} and \citet{Naz21}. Additionally, N-bearing molecules exhibit greater variation in relative abundances across different sources compared to O-bearing molecules. The relative abundances of most O-bearing molecules in G336.99-00.03 MM1 are consistent within an order of magnitude with those observed in the four hot cores of G9.62+0.19 and the Galactic centre hot cores SgrB2(N4) and SgrB2(N5). However, these abundances are 10 to 40 times lower than those observed in SgrB2(N2) and SgrB2(N3). Despite being in a hot core stage, G336.99-00.03 MM1 exhibits a richer diversity of N-bearing molecules compared to G9.62+0.19 MM4, MM7, MM8, and MM11. In contrast, the diversity and abundance of O-bearing molecules in G336.99-00.03 MM1 are comparable to those observed in these other hot cores, suggesting that their chemical evolution processes may share similar mechanisms. 

Given the challenges in accurately determining the column density of H$_{2}$ from dust continuum, we also discussed the abundance of the detected species from different sources relative to CH$_{3}$OH, which is one of the most abundant COMs in the ISM. The lower panel of Figure \ref{fig:fig. 9} shows the abundances with respect to CH$_{3}$OH. N-bearing species exhibit smaller variations for their abundances relative to CH$_{3}$OH compared to their H$_{2}$ normalized abundances, especially for Galactic center hot cores. Comparing the abundances of O-bearing species from the different sources shown in Figure \ref{fig:fig. 9}, it can be observed that some O-bearing molecules have similar abundance ratios (both with respect to H$_{2}$ and with respect to CH$_{3}$OH), suggesting that they may be chemically related.

The Pearson $r$ coefficient is computed in this work to assess possible connections between the observed molecular abundances of multiple species across the extensive sample of sources. The Pearson $r$ coefficient is calculated using the following equation:
\begin{equation}
r = \frac{\sum_{i=1}^{n} (X_{1,i} - \bar{X_1})(X_{2,i} - \bar{X_2})}{\sqrt{\sum_{i=1}^{n} (X_{1,i} - \bar{X_1})^2 \sum_{i=1}^{n} (X_{2,i} - \bar{X_2})^2}}
\label{eq:equation5}
\end{equation}
where $r$ is computed based on the observed abundances of two species, denoted as $X_{1}$ and $X_{2}$, while $\bar{X_1}$ and $\bar{X_2}$ represent their respective mean values. The resulting correlation matrix is then visualized as a heat map, where positive and negative correlation values indicate potential chemical relationships between species.

\begin{figure}[]
\centering
\includegraphics[width=0.8\columnwidth]{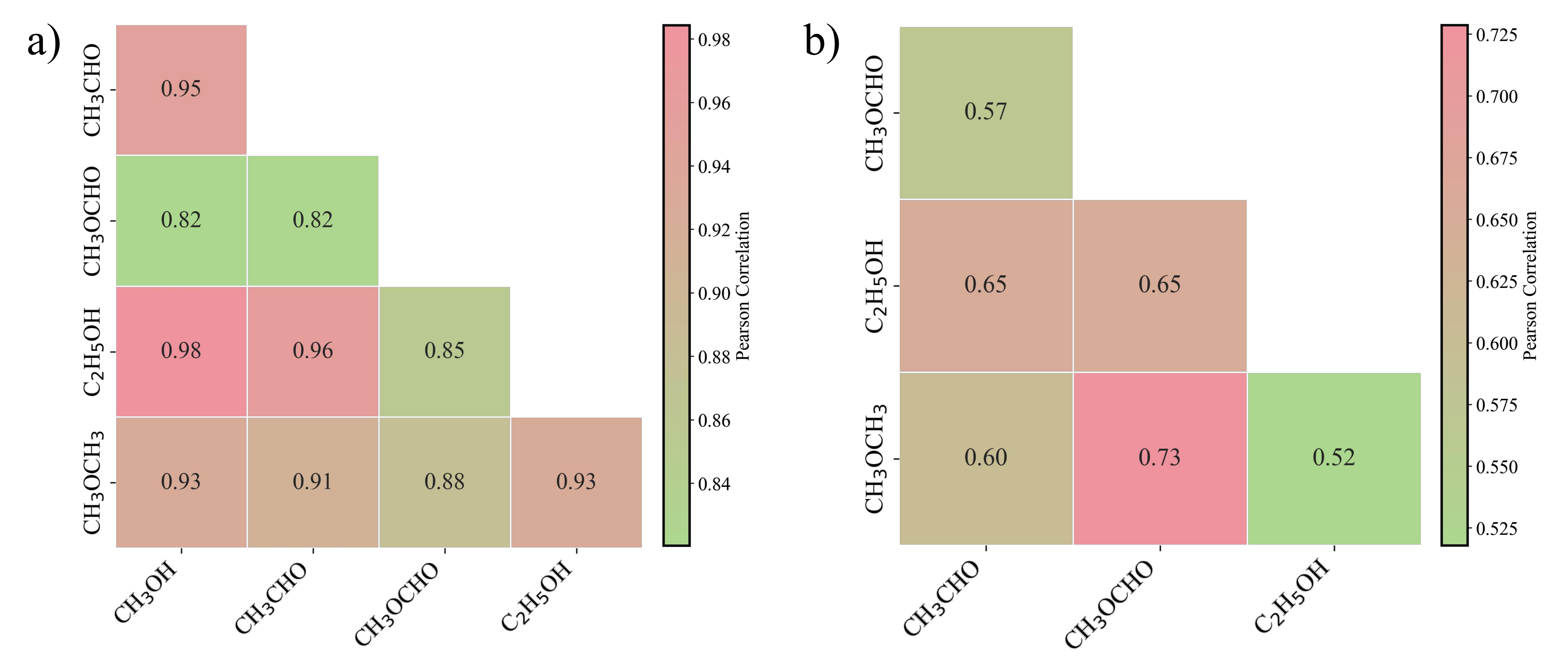}
\caption{(a) Heat maps showing the correlation of the observed molecular abundances with respect to H$_{2}$ of five O-bearing species: CH$_{3}$OH, CH$_{3}$CHO, CH$_{3}$OCHO, C$_{2}$H$_{5}$OH, and CH$_{3}$OCH$_{3}$. (b) The correlation of the observed molecular abundances with respect to CH$_{3}$OH of four O-bearing species: CH$_{3}$CHO, CH$_{3}$OCHO, C$_{2}$H$_{5}$OH, and CH$_{3}$OCH$_{3}$.The color scale represents the level of correlation, corresponds to the Pearson $r$ coefficient displayed in each box. Scatter plots for these correlations are presented in Figures \ref{fig:fig. 13} and \ref{fig:fig. 14}.
\label{fig:fig. 10}}
\end{figure}

Figure \ref{fig:fig. 10} (a) illustrates heat maps for the Pearson $r$ coefficient, showing the strong correlation of the observed molecular abundances relative to H$_{2}$ of five O-bearing species: CH$_{3}$OH, CH$_{3}$CHO, CH$_{3}$OCHO, C$_{2}$H$_{5}$OH and CH$_{3}$OCH$_{3}$. The color scale represents the level of correlation, with scatter plots provided in Figure \ref{fig:fig. 13}. Given the substantial uncertainties in estimating H$_{2}$ column density across different sources, the observed correlations of molecular abundances normalized by H$_{2}$ must be interpreted with caution. To mitigate this concern, we have presented an additional correlation using abundances normalized by CH$_{3}$OH (Figure \ref{fig:fig. 10} (b)). In this analysis, the correlations are significantly weaker, indicating that previously observed correlations might be driven by individual outliers or systematic uncertainties. Thus, while some positive correlations remain suggestive, these relationships require additional observational data to robustly establish potential chemical connections between these molecular species.

\subsubsection{Comparison between observed and modeled abundances} \label{sec:sec5.3.2}
\citet{Gar22} simulated a comprehensive three-phase (gas, grain, and ice mantle) chemical model to explore the formation and evolution of COMs in hot cores. This work represents one of the most comprehensive studies of this kind. This model begins with a free-fall collapse phase, during which the gas density rises from an initial n$_{\rm H}$ = $3 \times 10^{3}$ $\rm cm^{-3}$ to a final density of $2 \times 10^{8}$ $\rm cm^{-3}$, while maintaining a constant temperature of 10 K. This is followed by a warm-up phase, where the density remains fixed at $2 \times 10^{8}$ $\rm cm^{-3}$, and the temperature gradually increases from 8 K to 400 K. The warm-up phase is further divided into three scenarios based on timescales: Fast ($5 \times 10^{4}$ years), Medium ($2 \times 10^{5}$ years), and Slow ($1 \times 10^{6}$ years). The initial elemental abundances used in the model are $1.4 \times 10^{-4}$ for carbon, $7.5 \times 10^{-5}$ for nitrogen, $3.2 \times 10^{-4}$ for oxygen, and $8.0 \times 10^{-8}$ for sulfur, respectively. This section compares the molecular abundances relative to CH$_{3}$OH observed in G336.99-00.03 MM1 with predictions from the chemistry models presented by \citet{Gar22}. This comparison is physically reasonable due to the H$_{2}$ density of $6.6 \times 10^{6}$ $\rm cm^{-3}$ and dust temperature of 146 K in G336.99-00.03 MM1. However, we note the limitations of this comparison. Key parameters such as the cosmic-ray ionization rate $\zeta$ and visual extinction $A$$_{\rm V}$ remain unconstrained in our observations, which could influence the chemical evolution.

\begin{figure}[]
\centering
\includegraphics[width=0.6\columnwidth]{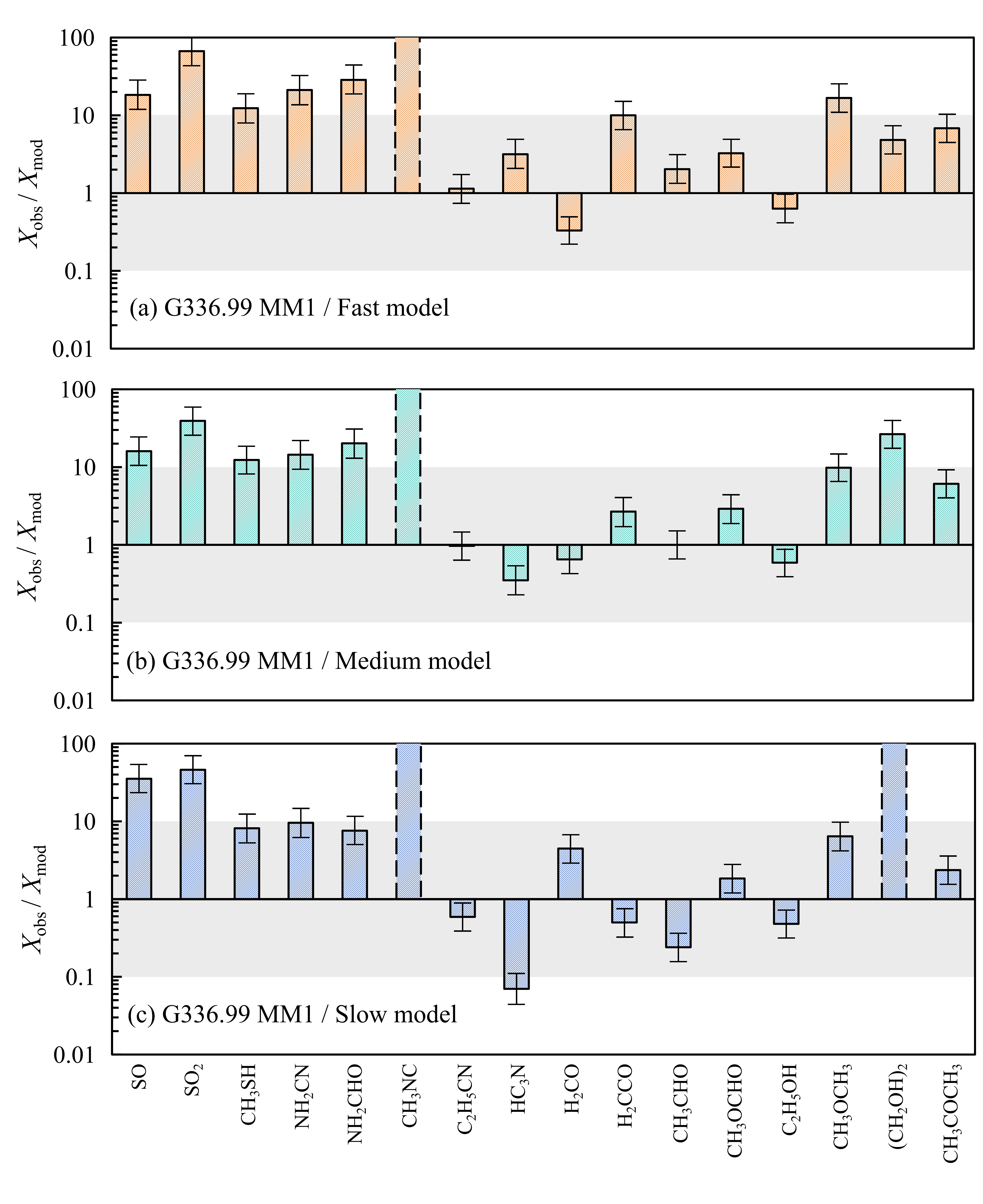}
\caption{Comparison between the observed molecular abundances relative to CH$_{3}$OH (X$_{obs}$) and the peak gas-phase molecular abundances predicted by models (X$_{mod}$), based on data from \citet{Gar22}. Bars represent the X$_{obs}$ / X$_{mod}$ ratio, with dashed lines indicating values that exceed the plot bounds. The shaded area highlights cases where the observed and modeled values differ by no more than 1 order of magnitude. (a) Peak gas-phase molecular abundances from the Fast warm-up models; (b) Peak gas-phase molecular abundances from the Medium warm-up models; (c) Peak gas-phase molecular abundances from the Slow warm-up models.
\label{fig:fig. 11}}
\end{figure}

Figure \ref{fig:fig. 11} compares the observed molecular abundances relative to CH$_{3}$OH with the peak gas-phase molecular abundances predicted by three warm-up timescale models: Fast (upper panel), Medium (middle panel), and Slow (lower panel). A 50\% error margin was assumed for the values derived from the chemical models. Overall, the observed abundances of most molecules align well with the model predictions, with some variations between the different warm-up timescales. The observed values for G336.99-00.03 MM1 are generally within the range produced by the models. However, discrepancies were noted for certain molecules. The CH$_{3}$NC production is severely deficient in all models, where the observed-to-modeled abundance ratio exceeds 100.

For S-bearing molecules, almost all models do not reproduce the observed abundances effectively, and only the Slow warm-up model performs slightly better, with a difference of less than 1 order of magnitude for CH$_{3}$SH. The performance for N-bearing molecules varies: the Slow warm-up model provides superior results for NH$_{2}$CN and NH$_{2}$CHO, but overproduces HC$_{3}$N. For O-bearing molecules, all models reproduce the abundance of most species. However, (CH$_{2}$OH)$_{2}$ in Slow warm-up models is severely underproduced.

A potential explanation for some of these discrepancies lies in the difference between the derived gas-phase temperatures of the detected molecules and the temperatures at which the molecules reach their peak abundances in the models. Figure \ref{fig:fig. 15} illustrates the derived gas-phase temperatures of the molecules detected in G336.99-00.03 MM1, alongside the temperatures corresponding to the peak gas-phase abundances predicted by the models. These peak temperatures result from the interplay of desorption from grain surfaces, gas-phase formation and destruction processes, and the underlying chemical assumptions in the models. It is evident that the observed molecular temperatures are consistently lower than those predicted by the models. It is worth noting that for the temperatures shown in Figure \ref{fig:fig. 15}, the slow model seems to reproduce the observations better. Despite some differences in abundance and temperature, the Slow warm-up timescale model proves to be the most suitable in reproducing the observed molecular abundances in G336.99-00.03 MM1 based on the currently analyzed molecules. Therefore, the following discussion will focus on the values derived from the Slow warm-up model.

\subsection{Possible formation pathways of several molecules} \label{sec:sec5.4}
\subsubsection{HC$_{3}$N and C$_{2}$H$_{5}$CN} \label{sec:sec5.4.1}
Previous observations have identified C$_{2}$H$_{5}$CN as a common N-bearing molecule in hot cores \citep{Bis07, Suz18, Pen22, Naz22}. Studies by \citet{Gar13} and \citet{Gar17, Gar22} suggest that the formation of C$_{2}$H$_{5}$CN is chemically linked to HC$_{3}$N through the following reactions:

\begin{equation}
\text{HC}_3\text{N} + \text{H} \rightarrow \text{C}_2\text{H}_2\text{CN} \ \text{(\rm grain surface)}
\label{eq:reaction1}
\end{equation}

\begin{equation}
\text{C}_2\text{H}_2\text{CN} + \text{H} \rightarrow \text{C}_2\text{H}_3\text{CN} \ \text{(\rm grain surface)}\label{eq:reaction2}
\end{equation}

\begin{equation}
\text{C}_2\text{H}_3\text{CN} + \text{H} \rightarrow \text{C}_2\text{H}_4\text{CN} \ \text{(\rm grain surface)}\label{eq:reaction3}
\end{equation}

\begin{equation}
\text{C}_2\text{H}_4\text{CN} + \text{H} \rightarrow \text{C}_2\text{H}_5\text{CN} \ \text{(\rm grain surface)}
\label{eq:reaction4}
\end{equation}

Reactions \ref{eq:reaction1} - \ref{eq:reaction4} indicate that C$_{2}$H$_{5}$CN is produced on grain surfaces via hydrogenation of HC$_{3}$N, a hypothesis also supported by \citet{Pen22}. For HC$_{3}$N, multiple chemical reactions in both the gas and ice phases are documented in astrochemical databases such as KIDA and UMIST \citep{Mce13, Wak15}. During the freefall collapse stage, \citet{Tan19} suggested that the following reactions significantly contribute to the formation of HC$_{3}$N:

\begin{equation}
\text{C}_4\text{H} + \text{N} \rightarrow \text{HC}_3\text{N} + \text{C} \ \text{(\rm grain surface)}
\label{eq:reaction5}
\end{equation}

\begin{equation}
\text{HC}_3\text{NH}^+ + \text{e}^- \rightarrow \text{HC}_3\text{N} + \text{H} \ \text{(\rm gas phase)}
\label{eq:reaction6}
\end{equation}

\citet{Has08, Cha09} suggested that HC$_{3}$N can be formed by neutral-neutral reaction between C$_{2}$H$_{2}$ and CN under hot core conditions:
\begin{equation}
\text{C}_2\text{H}_2 + \text{CN} \rightarrow \text{HC}_3\text{N} + \text{H} \ \text{(\rm grain surface)}
\label{eq:reaction7}
\end{equation}

Reaction \ref{eq:reaction7} was suggested as the main formation pathway of HC$_{3}$N based on observations of its three $^{13}$C isotopologues toward the high-mass star-forming region G28.28-0.36 \citep{Tan16}. While HC$_{3}$N can also form in the collapse phase, its abundance in cold cores is generally lower (typically around $\sim 10^{-9}$) compared to the higher abundances observed in hot cores. To investigate the formation mechanisms of HC$_{3}$N and C$_{2}$H$_{5}$CN in G336.99-00.03 MM1, we compared the estimated abundances of these molecules with the modeled values from \citet{Gar22}. The abundances of HC$_{3}$N and C$_{2}$H$_{5}$CN relative to CH$_{3}$OH in G336.99-00.03 MM1 were $(5.68 \pm 0.05) \times 10^{-3}$ and $(1.07 \pm 0.01) \times 10^{-2}$, respectively. These values are consistent with the modeled abundances within an order of magnitude. Based on this comparison and supported by previous chemical models, we suggest that C$_{2}$H$_{5}$CN in G336.99-00.03 MM1 may form on grain surfaces via hydrogenation of HC$_{3}$N, while HC$_{3}$N itself could be produced through Reaction \ref{eq:reaction7}. However, we emphasize that these proposed mechanisms remain speculative since such agreement does not provide definitive proof of specific reaction pathways and would require further observational, experimental, or theoretical studies to confirm.

\subsubsection{CH$_{3}$SH and CH$_{3}$OH} \label{sec:sec5.4.2}

CH$_{3}$OH is well established as being primarily formed on grain surfaces through hydrogenation of CO \citep{Wat02, Fuc09}. Similarly, CH$_{3}$SH shares a structural and chemical analogy with CH$_{3}$OH, as the sulfur atom in CH$_{3}$SH replaces the oxygen atom in CH$_{3}$OH, within the molecular framework, though their formation pathways are distinct. The key formation pathways of CH$_{3}$SH and CH$_{3}$OH are as follows:
\begin{equation}
\text{CS} \rightarrow \text{HCS} \rightarrow \text{H}_2\text{CS} \rightarrow \text{CH}_3\text{S} \rightarrow \text{CH}_3\text{SH} \ \text{(\rm grain surface)}
\label{eq:reaction8}
\end{equation}
\begin{equation}
\text{CO} \rightarrow \text{HCO} \rightarrow \text{H}_2\text{CO} \rightarrow \text{CH}_3\text{O} \rightarrow \text{CH}_3\text{OH} \ \text{(\rm grain surface)}
\label{eq:reaction9}
\end{equation}

In G336.99-00.03 MM1, the observed abundance relative to CH$_{3}$OH of CH$_{3}$SH is $3.6 \times 10^{-3}$, which closely matches the modeled abundance predicted by \citet{Gar22}. Previous studies \citep{Maj16, Mul16, Vid17} have proposed that CH$_{3}$SH could form through the hydrogenation of CS on grain surfaces. The agreement between the observed and modeled abundances suggests that the formation of CH$_{3}$SH in G336.99-00.03 MM1 may involve grain-surface chemistry, but it does not definitively confirm the specific reaction pathways. Additional observational and experimental evidence is needed to conclusively determine the formation mechanism of CH$_{3}$SH in this source.

\subsubsection{NH$_{2}$CN and NH$_{2}$CHO} \label{sec:sec5.4.3}
It has been proposed that the neutral-neutral reaction between NH$_{2}$ and CN radicals is the most efficient pathway for producing NH$_{2}$CN on grain surfaces in hot cores and hot corinos \citep{Cou18, Zha23}, as the following:
\begin{equation}
\text{NH}_2 + \text{CN} \rightarrow \text{NH}_2\text{CN} \ \text{(\rm grain surface)}
\label{eq:reaction10}
\end{equation}

Given that NH$_{2}$CHO and NH$_{2}$CN share the same functional group, this suggests a possible chemical link between these two molecules. Initially, \citet{Qua07} claimed that NH$_{2}$CHO could be formed via ion-molecule reactions between $\text{NH}_4^+$ and H$_{2}$CO, followed by electron recombination. Subsequently, \citet{Gar08}, \citet{Gar13} and \citet{Bar15} proposed that NH$_{2}$CHO can also form in the gas phase through a barrierless reaction between NH$_{2}$ and H$_{2}$CO:
\begin{equation}
\text{NH}_2 + \text{H}_2\text{CO} \rightarrow \text{NH}_2\text{CHO} + \text{H} \ \text{(\rm gas phase)}
\label{eq:reaction11}
\end{equation}

Numerous observations suggest that Reaction \ref{eq:reaction11} may be a major formation pathway for NH$_{2}$CHO in the gas phase, including detections toward the hot corino IRAS 16293-2422B, the shock region L1157-B1, hot cores such as SgrB2 (N), G10.47+0.03, Orion KL, and G31.41+0.31 \citep{Hal11, Kah13, Cou16, Cod17, Gor20, Col21}. However, recent laboratory and computational studies by \citet{Dou22} indicate that Reaction \ref{eq:reaction11} is not a significant source of NH$_{2}$CHO in interstellar environments. Instead, alternative pathways involving reactions that occur on the surface of dust grains (e.g., NH$_{2}$ + HCO), may play a more dominant role in the formation of NH$_{2}$CHO \citep{Que18, Rim18}. In G336.99-00.03 MM1, the observed abundances relative to CH$_{3}$OH of NH$_{2}$CN and NH$_{2}$CHO are $5.8 \times 10^{-4}$ and $4.9 \times 10^{-2}$, respectively. These values align reasonably well with the modeled abundances from \citet{Gar22}, whose models incorporate both gas-phase and grain-surface chemistry. This agreement indicates that grain-surface reactions, rather than purely gas-phase processes, likely play a key role in the formation of NH$_{2}$CHO in G336.99-00.03 MM1. Furthermore, these reactions support the idea of a chemical link between NH$_{2}$CN and NH$_{2}$CHO, as both molecules share NH$_{2}$ as a common precursor.

\subsubsection{C$_{2}$H$_{n}$O family} \label{sec:sec5.4.4}

\citet{Chu21} proposed that C$_{2}$H$_{n}$O-type COMs (n = 2, 4, and 6) may be chemically related. However, observations by \citet{Pen22} found no clear relationship between the column densities of H$_{2}$CCO and other C$_{2}$H$_{n}$O molecules, challenging this hypothesis. \citet{Tan23} further pointed out that using column densities in correlation analyses can be misleading due to the confounding effect of total gas column density ($\mathit{N}_{{\rm H}_{2}}$). They recommended using molecular abundances relative to H$_{2}$ as a more reliable metric. In our work, we adopted this approach to explore possible correlations among C$_{2}$H$_{n}$O species. As shown in Figure \ref{fig:fig. 10} (a), their H$_{2}$-normalized abundances exhibit strong positive correlation, of C$_{2}$H$_{n}$O molecules, which become less pronounced when normalized to CH$_{3}$OH.

\citet{Gar22} proposed a formation pathway for H$_{2}$CCO on grain surfaces via reaction of CH$_{2}$ with CO:
\begin{equation}
\text{CH}_2 + \text{CO} \rightarrow \text{H}_2\text{CCO} \ \text{(\rm grain surface)}
\label{eq:reaction12}
\end{equation}

Under cold conditions, H$_{2}$CCO is hydrogenated to form CH$_{3}$CHO through the following reactions:
\begin{equation}
\text{H}_2\text{CCO} + \text{H} \rightarrow \text{CH}_3\text{CO} \ \text{(\rm grain surface)}
\label{eq:reaction13}
\end{equation}
\begin{equation}
\text{CH}_3\text{CO} + \text{H} \rightarrow \text{CH}_3\text{CHO} \ \text{(\rm grain surface)}
\label{eq:reaction14}
\end{equation}

If this pathway is indeed effective, H$_{2}$CCO and CH$_{3}$CHO are expected to be chemically linked. The Pearson $r$ coefficient (calculated using Equation \ref{eq:equation5}) between their H$_{2}$-normalized abundances is 0.94, while the CH$_{3}$OH-normalized value drops to 0.28. This discrepancy, along with the limited sample size (H$_{2}$CCO was detected in only six sources), suggests that the correlation should be interpreted with caution and warrants further investigation.

For C$_{2}$H$_{5}$OH and CH$_{3}$OCH$_{3}$ in the C$_{2}$H$_{n}$O family, \citet{Mai15} proposed that their formation may be driven by radical-radical reactions:
\begin{equation}
\text{CH}_3 + \text{CH}_2\text{OH} \rightarrow \text{C}_2\text{H}_5\text{OH} \ \text{(\rm grain surface)}
\label{eq:reaction15}
\end{equation}
\begin{equation}
\text{CH}_3 + \text{CH}_3\text{O} \rightarrow \text{CH}_3\text{OCH}_3 \ \text{(\rm grain surface)}
\label{eq:reaction16}
\end{equation}

Both CH$_{2}$OH and CH$_{3}$O are produced through the photolysis of CH$_{3}$OH \citep{Ben07a, Ben07b}:
\begin{equation}
\text{CH}_3\text{OH} + h\nu \rightarrow \text{CH}_2\text{OH} \ \text{(\rm grain surface)}
\label{eq:reaction17}
\end{equation}
\begin{equation}
\text{CH}_3\text{OH} + h\nu \rightarrow \text{CH}_3\text{O} \ \text{(\rm grain surface)}
\label{eq:reaction18}
\end{equation}

If these formation pathways are effective, CH$_{3}$OH serves as a precursor for C$_{2}$H$_{5}$OH and CH$_{3}$OCH$_{3}$, leading to positive correlations. Furthermore, numerous observations corroborate the tight correlation of C$_{2}$H$_{5}$OH and CH$_{3}$OCH$_{3}$ with CH$_{3}$OH \citep{Bis07, Law21, Pen22}.

\section{Conclusions} \label{sec:sec6}
In this study, a line survey was conducted towards the high-mass star forming region G336.99-00.03 using the ALMA interferometer at Band 3, covering a frequency range of 97.5–101.4 GHz. The main findings are summarized as follows:
   \begin{enumerate}
      \item \textbf{Molecular Inventory in MM1 and MM2:}
      Approximately 300 emission lines were identified in MM1, attributed to 19 molecular species, 8 isotopologues and several vibrationally excited states of HC$_{3}$N, C$_{2}$H$_{5}$CN, as well as torsionally excited states of CH$_{3}$OH. In contrast, only 7 emission lines corresponding to 5 species were detected in MM2. Through spectral modeling under the assumption of LTE, excitation temperatures (T$_{ex}$) ranging from 73 to 249 K and molecular column densities (N$_{T}$) in the range of 4.42$\times$10$^{14}$ – 4.00$\times$10$^{18}$ cm$^{-2}$ were derived.
      
      \item \textbf{Isotopic Ratios:}
      The isotopic ratios of $^{12}$C/$^{13}$C and $^{32}$S/$^{34}$S in G336.99-00.03 MM1 are generally consistent with Galactic trends, except for HC$_{3}$N. In this case, the deviation likely results from an underestimated opacity in the $\nu$=0 transition, as indicated by the $\nu$$_{7}$=1 line. The $^{16}$O/$^{18}$O ratio also shows a notable deviation from expected values, suggesting further study with a larger source sample is needed.
            
      \item \textbf{Evolutionary Stages of MM1 and MM2:}
      The evolutionary stages of the two millimeter continuum sources in G336.99-00.03 were examined. MM1, characterized by a rich molecular inventory, is classified as being in a HC phase. In contrast, MM2, where hydrogen recombination lines such as H$_{40 \alpha}$ and H$_{50 \beta}$ emissions were detected, is considered as an HII stage.
      
      \item \textbf{Comparison with Other Sources:}
      Molecular abundances relative to H$_{2}$ and CH$_{3}$OH in G336.99-00.03 MM1 were compared with those from other sources and chemical models. Variations among sources were more pronounced for N-bearing molecules than for O-bearing molecules. O-bearing molecules showed a stronger correlation when abundances were normalized to H$_{2}$, indicating potential chemical links. However, this correlation weakened when normalized to CH$_{3}$OH. The observed abundances in MM1 show partial agreement with chemical model predictions for three different warm-up timescales (Fast, Medium, and Slow). The Slow warm-up model provides the closest match overall, but 5 species still deviate by more than an order of magnitude.

      This work provides valuable insight into the chemical complexity and evolutionary stages of high-mass star-forming regions. The results, particularly the comparison with chemical models, provide important constraints for astrochemical simulations and highlight the role of hot cores in the development of chemical complexity in the interstellar medium.

   \end{enumerate}

\begin{acknowledgments}
This work has been supported by the Fundamental Research Funds for the Central Universities (Grant Nos. 2024CDJGF-025 and 2023CDJXY-045), Chongqing Municipal Natural Science Foundation General Program (Grant No. cstc2021jcyj-msxmX0867), National Natural Science Foundation of China (Grant Nos. 12103010, 12041305, and 12033005), the Tianchi Talent Program of Xinjiang Uygur Autonomous Region, the China-Chile Joint Research Fund (CCJRF, No. 2211), and the High-Performance Computing Platform of Peking University. This work makes use of the following ALMA data: ADS/JAO.ALMA$\#$2019.1.00685.S. ALMA is a partnership of ESO (representing its member states), NSF (USA), and NINS (Japan), together with NRC (Canada), MOST and ASIAA (Taiwan, China), and KASI (Republic of Korea), in cooperation with the Republic of Chile. The Joint ALMA Observatory is operated by ESO, AUI/NRAO, and NAOJ. 
\end{acknowledgments}

%

\vspace{5mm}
\facilities{ALMA}


\software{CASA \citep{Mcm07},  
          GILDAS ({\url{http://www.iram.fr/IRAMFR/GILDAS}}), 
          CARTA \citep{Com21}
          }


\clearpage

\appendix

\section{APPENDIX A}

\begin{figure}[h]
\plotone{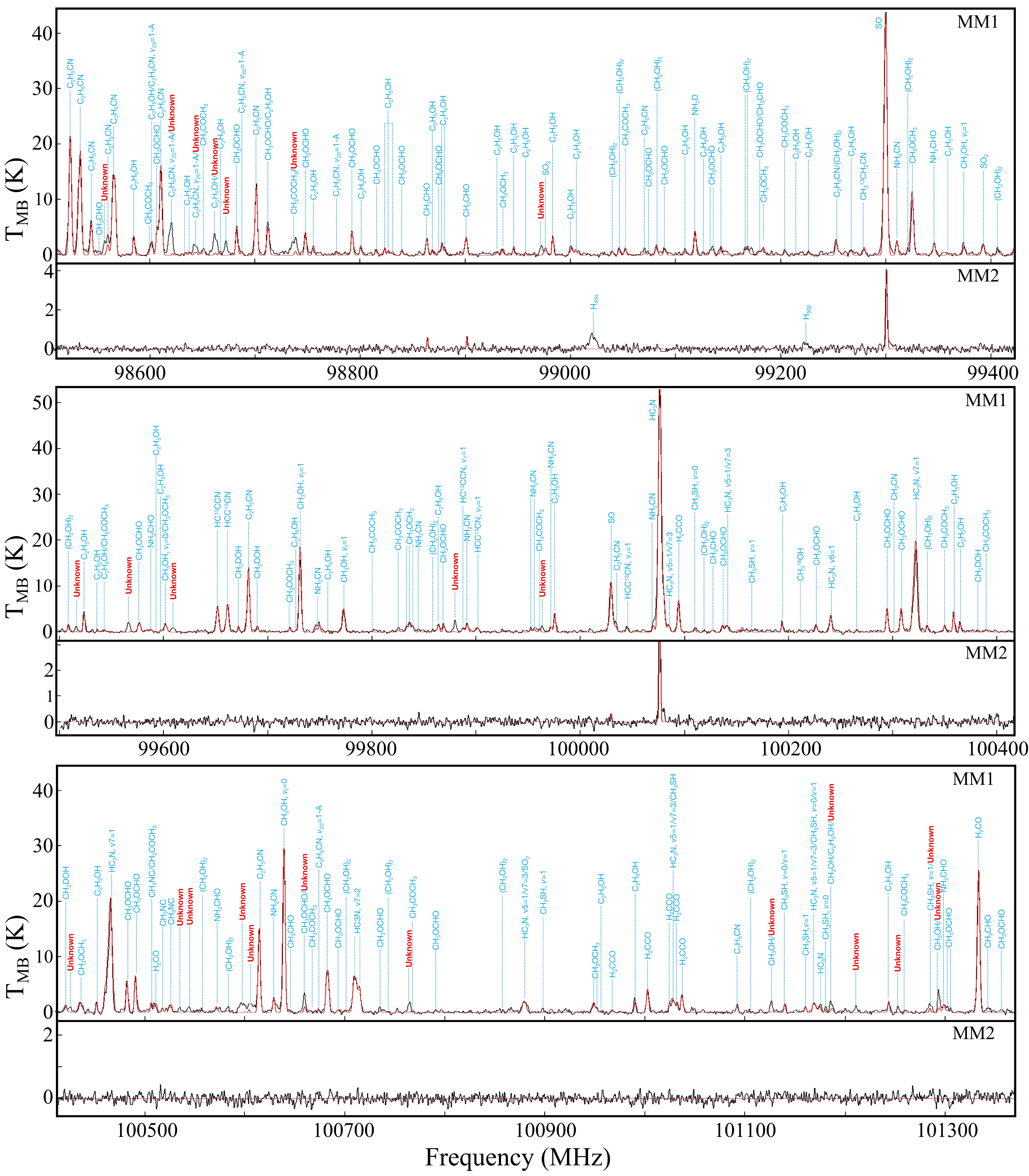}
\caption{Observed and modeled synthetic spectra for MM1 and MM2. The black solid lines represent the observed spectra, while the red solid lines correspond to the modeled spectra. Unidentified lines are marked in red and labeled as “Unknown”.
\label{fig:fig. 12}}
\end{figure}

The emission lines identified in G336.99-00.03 are shown in Figure \ref{fig:fig. 12}

\clearpage

\section{APPENDIX B}

\begin{deluxetable*}{cccccc}[htp]
\tablenum{5}
\tablecaption{Molecular Line Parameters. \label{table:table B1}}
\tablewidth{0pt}
\tablehead{
  \colhead{Species} & \colhead{Freq [MHz]} & \colhead{$E_{\text{up}}$ [K]} & \colhead{$A_{ij}$ [s$^{-1}$]} & \colhead{Transitions} & \colhead{$\tau$}
}
\startdata
HC$_{3}$N, $\nu$=0       & 100076.392 & 28.8  & $7.77 \times 10^{-5}$ & 11 -- 10          & 1.05 \\
\hline
CH$_{3}$OH, $\nu_{t}$=0    & 97582.798  & 21.6  & $2.63 \times 10^{-6}$ & 2 1 1 -- 1 1 0    & 0.61 \\
                  & 97677.684  & 729.3 & $1.44 \times 10^{-6}$ & 21 616 -- 22 517  & 0.04 \\
                  & 97678.803  & 729.3 & $1.44 \times 10^{-6}$ & 21 615 -- 22 518  & 0.04 \\
                  & 98030.648  & 889.0 & $1.43 \times 10^{-6}$ & 24 619 -- 23 716  & 0.02 \\
                  & 98030.686  & 889.0 & $1.43 \times 10^{-6}$ & 24 618 -- 23 717  & 0.02 \\
                  & 100638.872 & 233.6 & $1.69 \times 10^{-6}$ & 13 212 -- 12 310  & 0.54 \\
\hline
C$_{2}$H$_{5}$CN, $\nu$=0     & 98177.574  & 32.8  & $7.55 \times 10^{-5}$ & 11 210 -- 10 2 9  & 0.19 \\
                  & 98523.872  & 68.4  & $5.54 \times 10^{-5}$ & 11 6 5 -- 10 6 4  & 0.09 \\
                  & 98524.672  & 82.8  & $4.70 \times 10^{-5}$ & 11 7 4 -- 10 7 3  & 0.07 \\
                  & 98532.084  & 99.5  & $3.72 \times 10^{-5}$ & 11 8 3 -- 10 8 2  & 0.05 \\
                  & 98533.987  & 56.2  & $6.26 \times 10^{-5}$ & 11 5 6 -- 10 5 5  & 0.13 \\
                  & 98544.164  & 118.3 & $2.61 \times 10^{-5}$ & 11 9 3 -- 10 9 2  & 0.03 \\
                  & 98566.792  & 46.2  & $6.86 \times 10^{-5}$ & 11 4 7 -- 10 4 6  & 0.15 \\
                  & 98701.101  & 38.4  & $7.35 \times 10^{-5}$ & 11 3 8 -- 10 3 7  & 0.17 \\
                  & 99681.461  & 33.0  & $7.90 \times 10^{-5}$ & 11 2 9 -- 10 2 8  & 0.19 \\
                  & 100614.281 & 30.1  & $8.33 \times 10^{-5}$ & 11 110 -- 10 1 9  & 0.20 \\
\enddata
\end{deluxetable*}

\clearpage

\section{APPENDIX C}
Molecular abundances relative to H$_{2}$ and CH$_3$OH derived in G336.99-00.03 MM1, compared with values reported for other sources and models. This comparison provides insights into chemical diversity and evolutionary differences among high-mass star-forming regions.

\begin{deluxetable*}{cccccccccc}[h]
\tablenum{6}
\tablecaption{Molecular abundances relative to H$_{2}$ in different hot cores and models. \label{table:table C1}}
\tablewidth{0pt}
\tablehead{
\colhead{Sources} & \colhead{CH$_{3}$SH} & \colhead{NH$_{2}$CN} & \colhead{NH$_{2}$CHO} &
\colhead{CH$_{3}$NC} & \colhead{C$_{2}$H$_{5}$CN} & \colhead{HC$_{3}$N} & \colhead{HC$_{5}$N} &
\colhead{H$_{2}$CCO} & \colhead{Ref}
}
\startdata
G336MM1     & 7.99(-9)  & 1.29(-9)   & 1.08(-7)  & 5.39(-10)  & 2.39(-8)  & 1.27(-8)  & 2.47(-10)  & 1.68(-8)  &  This work      \\
G9.62MM7    & -         & -          & -         & -          & 1.90(-8)  & 3.47(-9)  & 1.30(-9)   & 7.00(-9)  &  (1)  \\
G9.62MM8    & 9.70(-8)  & -          & -         & -          & 9.00(-8)  & 6.19(-9)  & 7.00(-10)  & 1.50(-8)  &  (1)  \\
G9.62MM4    & -         & -          & -         & -          & 1.30(-8)  & 4.45(-9)  & 1.80(-9)   & 1.50(-8)  &  (1)  \\
G9.62MM11   & -         & -          & -         & -          & -         & -         & -          & 6.00(-9)  &  (1)  \\
SgrB2(N2)   & 2.10(-7)  & -          & 2.12(-6)  & 6.06(-9)   & 3.80(-6)  & 6.39(-7)  & -          & 2.00(-7)  &  (2)(3)(4)(5)(6)  \\
SgrB2(N3)   & 9.80(-8)  & -          & 3.89(-8)  & -          & 3.60(-7)  & -         & -          & -         &  (7)  \\
SgrB2(N4)   & 3.70(-9)  & -          & 3.40(-9)  & -          & 5.50(-9)  & -         & -          & -         &  (7)  \\
SgrB2(N5)   & 3.90(-8)  & -          & 1.97(-8)  & -          & 7.60(-8)  & -         & -          & -         &  (7)  \\
Model-Fast  & 3.10(-9)  & 4.10(-10)  & 1.90(-8)  & 3.00(-12)  & 1.00(-7)  & 2.00(-8)  & -          & 8.10(-9)  &  (8)  \\
Model-Medium& 3.10(-9)  & 4.00(-10)  & 2.50(-8)  & 1.90(-12)  & 1.10(-7)  & 1.70(-7)  & -          & 2.90(-8)  &  (8)  \\
Model-Slow  & 3.70(-9)  & 5.00(-10)  & 5.30(-8)  & 7.50(-12)  & 1.50(-7)  & 6.70(-7)  & -          & 1.20(-7)  &  (8)  \\
\hline
Sources     & CH$_3$OH  & CH$_3$CHO  & CH$_3$OCHO & C$_2$H$_5$OH & CH$_3$OCH$_3$ & (CH$_2$OH)$_2$  & CH$_3$COCH$_3$ & -  & Ref\\
\hline                                                                                             
G336MM1     & 2.23(-6)  & 1.63(-8)   & 8.88(-8)   & 8.55(-8)     & 3.73(-7)      & 5.92(-8)        & 8.99(-9)       & -  &  This work      \\
G9.62MM7    & 1.30(-6)  & 1.50(-8)   & 4.80(-8)   & 4.80(-8)     & 7.80(-8)      & -               & -              & -  &  (1)  \\
G9.62MM8    & 1.05(-6)  & 1.30(-8)   & 1.55(-7)   & 5.50(-8)     & 1.26(-7)      & -               & 2.60(-8)       & -  &  (1)  \\
G9.62MM4    & 2.27(-6)  & 1.60(-8)   & 3.94(-7)   & 1.39(-7)     & 2.00(-7)      & -               & 3.30(-8)       & -  &  (1)  \\
G9.62MM11   & 7.14(-7)  & 1.40(-8)   & 1.11(-7)   & 3.20(-8)     & 1.04(-7)      & -               & -              & -  &  (1)  \\
SgrB2(N2)   & 2.40(-5)  & 2.60(-7)   & 7.30(-7)   & 1.20(-6)     & 1.30(-6)      & -               & 2.82(-7)      & -  &  (2)(3)(4)(5)(6)  \\
SgrB2(N3)   & 8.30(-6)  & 9.40(-8)   & 1.90(-6)   & 3.40(-7)     & 1.10(-6)      & -               & -              & -  &  (7)  \\
SgrB2(N4)   & 9.80(-8)  & 3.90(-9)   & 4.30(-8)   & 9.80(-9)     & 3.50(-8)      & -               & -              & -  &  (7)  \\
SgrB2(N5)   & 2.20(-6)  & 2.80(-8)   & 2.70(-7)   & 1.10(-7)     & 5.00(-7)      & -               & -              & -  &  (7)  \\
Model-Fast  & 1.10(-5)  & 3.90(-8)   & 1.90(-7)   & 6.60(-7)     & 1.10(-7)      & 6.00(-8)        & 6.40(-9)       & -  &  (8)  \\
Model-Medium& 1.00(-5)  & 7.60(-8)   & 2.00(-7)   & 6.80(-7)     & 1.80(-7)      & 1.10(-8)        & 7.00(-9)       & -  &  (8)  \\
Model-Slow  & 8.30(-6)  & 2.50(-7)   & 2.50(-7)   & 6.60(-7)     & 2.20(-7)      & 2.20(-12)       & 1.40(-8)       & -  &  (8)  \\
\enddata
\tablecomments{a(b) = a$\times$10$^{\rm b}$. Observational data are from: (1)\citet{Pen22}; (2)\citet{Bel16}; (3)\citet{Bel17}; (4)\citet{Wil20}; (5)\citet{Ord19};  (6)\citet{Jor20}; (7)\citet{Bon19}. Model predictions are from: (8)\citet{Gar22}.
}
\end{deluxetable*}
\clearpage

\begin{deluxetable*}{ccccccccc}[h]
\tablenum{7}
\tablecaption{Molecular abundances relative to CH$_3$OH in different hot cores and models. \label{table:table C2}}
\tablewidth{0pt}
\tablehead{
\colhead{Sources} & \colhead{CH$_{3}$SH} & \colhead{NH$_{2}$CN} & \colhead{NH$_{2}$CHO} &
\colhead{CH$_{3}$NC} & \colhead{C$_{2}$H$_{5}$CN} & \colhead{HC$_{3}$N} & \colhead{HC$_{5}$N} &
\colhead{Ref}
}
\startdata
G336MM1     & 3.58(-3)  & 5.77(-4)  & 4.85(-2)  & 2.41(-4)  & 1.07(-2)  & 5.68(-3)  & 1.11(-4)     &  This work      \\
G9.62MM7    & -         & -         & -         & -         & 1.47(-2)  & 2.80(-3)  & 1.00(-3)     &  (1)  \\
G9.62MM8    & 9.25(-2)  & -         & -         & -         & 8.38(-2)  & 5.94(-3)  & 6.50(-4)     &  (1)  \\
G9.62MM4    & -         & -         & -         & -         & 5.73(-3)  & 1.91(-3)  & 8.00(-4)     &  (1)  \\
G9.62MM11   & -         & -         & -         & -         & -         & -         & -            &  (1)  \\
SgrB2(N2)   & 8.75(-3)  & -         & 8.84(-2)  & 2.53(-4)  & 1.58(-1)  & 2.66(-2)  & -            &  (2)(3)(4)(5)(6)  \\
SgrB2(N3)   & 1.18(-2)  & -         & 4.69(-3)  & -         & 4.34(-2)  & -         & -            &  (7)  \\
SgrB2(N4)   & 3.78(-2)  & -         & 3.47(-2)  & -         & 5.61(-2)  & -         & -            &  (7)  \\
SgrB2(N5)   & 1.77(-2)  & -         & 8.95(-3)  & -         & 3.45(-2)  & -         & -            &  (7)  \\
Model-Fast  & 2.90(-4)  & 3.73(-5)  & 1.70(-3)  & 2.80(-7)  & 9.40(-3)  & 1.80(-3)  & -            &  (8)  \\
Model-Medium& 2.90(-4)  & 4.00(-5)  & 2.40(-3)  & 1.80(-7)  & 1.10(-2)  & 1.60(-2)  & -            &  (8)  \\
Model-Slow  & 4.40(-4)  & 6.02(-5)  & 6.40(-3)  & 9.10(-8)  & 1.80(-2)  & 8.10(-2)  & -            &  (8)  \\
\hline
Sources       & H$_2$CCO  & CH$_3$CHO  & CH$_3$OCHO & C$_2$H$_5$OH & CH$_3$OCH$_3$ & (CH$_2$OH)$_2$  & CH$_3$COCH$_3$ &  Ref  \\
\hline                                                                                   
G336MM1       & 7.50(-3)  & 7.30(-3)   & 3.98(-2)   & 3.82(-2)     & 1.67(-1)      & 2.65(-2)      & 4.02(-3)  &  This work      \\
G9.62MM7      & 5.33(-3)  & 1.13(-2)   & 3.67(-2)   & 3.67(-2)     & 6.00(-2)      & -             & -         &  (1)  \\
G9.62MM8      & 1.38(-2)  & 1.25(-2)   & 1.48(-1)   & 5.25(-2)     & 1.20(-1)      & -             & 2.50(-2)  &  (1)  \\
G9.62MM4      & 6.67(-3)  & 7.07(-3)   & 1.73(-1)   & 6.13(-2)     & 8.80(-2)      & -             & 1.47(-2)  &  (1)  \\
G9.62MM11     & 8.00(-3)  & 1.90(-2)   & 1.55(-1)   & 4.50(-2)     & 1.45(-1)      & -             & -         &  (1)  \\
SgrB2(N2)     & 8.33(-3)  & 1.08(-2)   & 3.04(-2)   & 5.00(-2)     & 5.42(-2)      & -             & 1.18(-2)  &  (2)(3)(4)(5)(6)  \\
SgrB2(N3)     & -         & 1.13(-2)   & 2.29(-1)   & 4.10(-2)     & 1.33(-1)      & -             & -         &  (7)  \\
SgrB2(N4)     & -         & 3.98(-2)   & 4.39(-1)   & 1.00(-1)     & 3.57(-1)      & -             & -         &  (7)  \\
SgrB2(N5)     & -         & 1.27(-2)   & 1.23(-1)   & 5.00(-2)     & 2.27(-1)      & -             & -         &  (7)  \\
Model-Fast    & 7.50(-4)  & 3.60(-3)   & 1.70(-2)   & 6.10(-2)     & 1.00(-2)      & 5.50(-3)      & 3.30(-4)  &  (8)  \\
Model-Medium  & 2.80(-3)  & 7.30(-3)   & 1.90(-2)   & 6.50(-2)     & 1.70(-2)      & 1.00(-3)      & 3.40(-4)  &  (8)  \\
Model-Slow    & 1.50(-2)  & 3.00(-2)   & 3.00(-2)   & 8.00(-2)     & 2.60(-2)      & 2.70(-7)      & 5.00(-4)  &  (8)  \\
\enddata
\tablecomments{a(b) = a$\times$10$^{\rm b}$. Observational data are from: (1)\citet{Pen22}; (2)\citet{Bel16}; (3)\citet{Bel17}; (4)\citet{Wil20}; (5)\citet{Ord19};  (6)\citet{Jor20}; (7)\citet{Bon19}. Model predictions are from: (8)\citet{Gar22}.
}
\end{deluxetable*}
\clearpage

\section{APPENDIX D}
Correlation analysis of the relative abundances of O-bearing molecules. This appendix presents detailed scatter plots and statistical correlations to explore potential chemical links among O-bearing species in G336.99-00.03 MM1 and other sources.

\begin{figure}[h]
\plotone{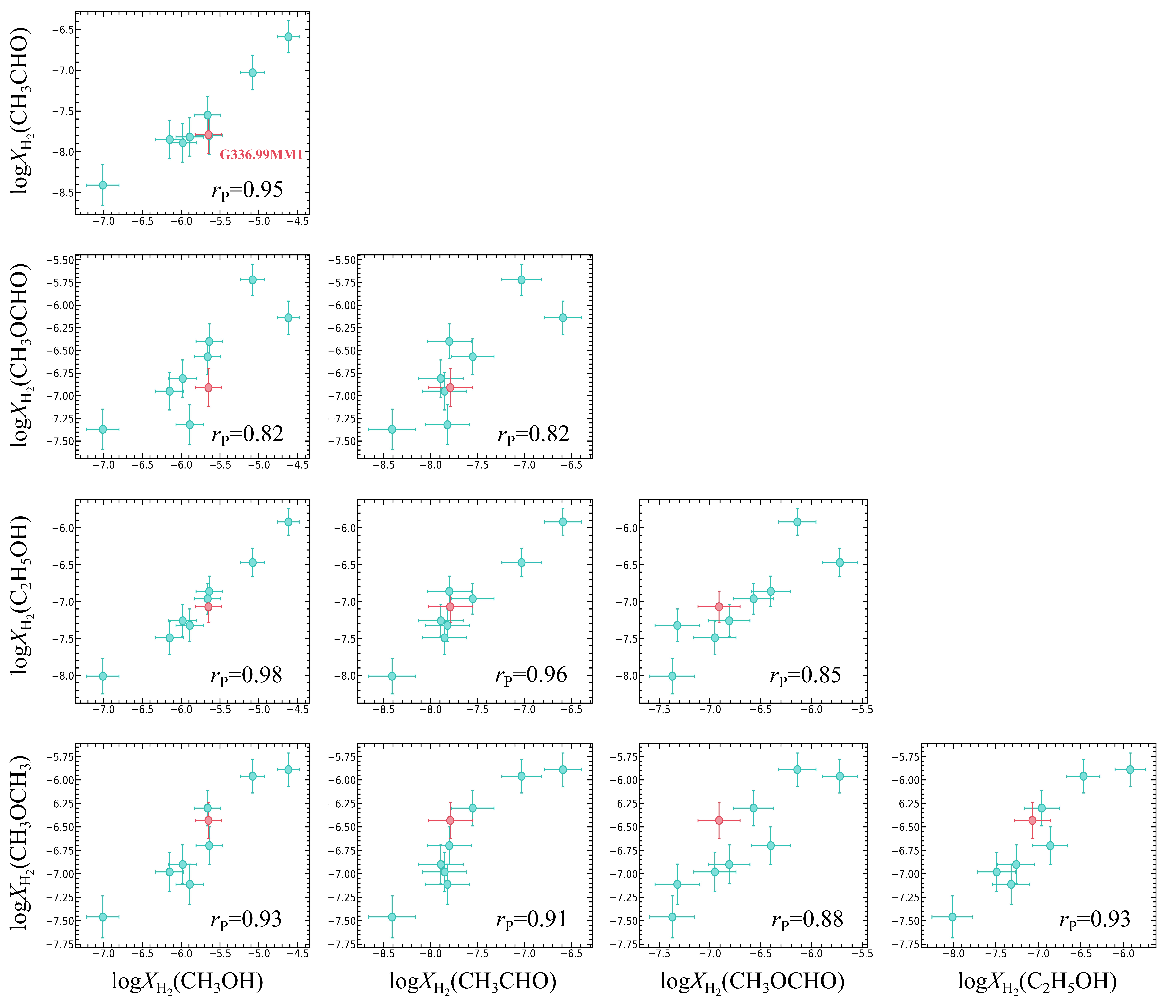}
\caption{Scatter plots of molecular abundances with respect to H$_{2}$, arranged consistently with the layout of the panel (a) of Figure \ref{fig:fig. 10}. The vertical error bars indicate a 20\% uncertainty in the values. The Pearson correlation coefficient ($r$) is given at the bottom right corner of each panel.
\label{fig:fig. 13}}
\end{figure}

\begin{figure}[h]
\plotone{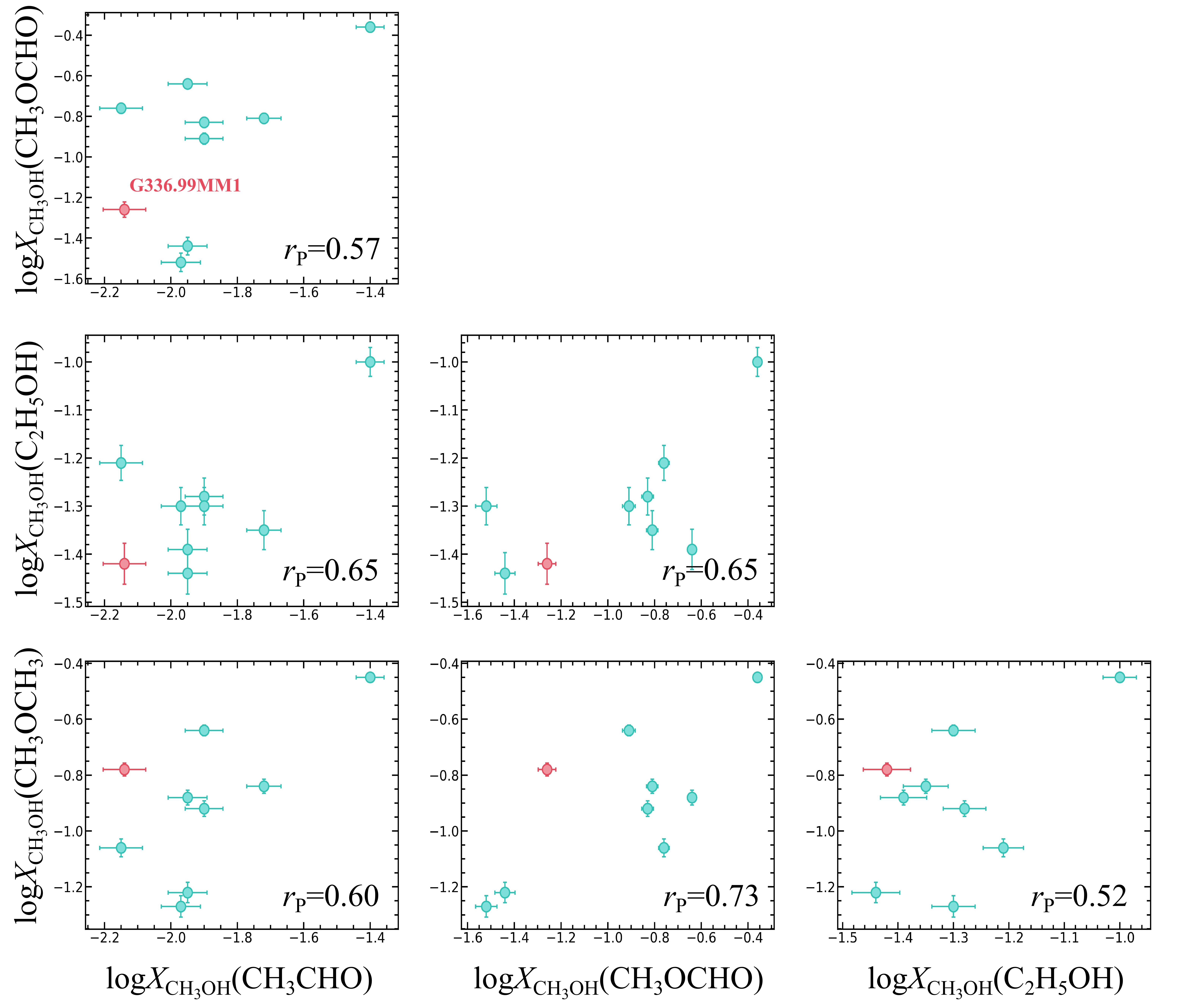}
\caption{Scatter plots of molecular abundances with respect to CH$_{3}$OH, arranged consistently with the layout of the panel (b) of Figure \ref{fig:fig. 10}. The vertical error bars indicate a 20\% uncertainty in the values. The Pearson correlation coefficient ($r$) is given at the bottom right corner of each panel.
\label{fig:fig. 14}}
\end{figure}

\clearpage

\section{APPENDIX E}
Comparison of the gas-phase temperatures of molecules detected in G336.99-00.03 MM1 with those predicted by chemical models. This appendix highlights temperature discrepancies and their potential impact on the agreement between observed and modeled molecular abundances.

\begin{figure}[h]
\plotone{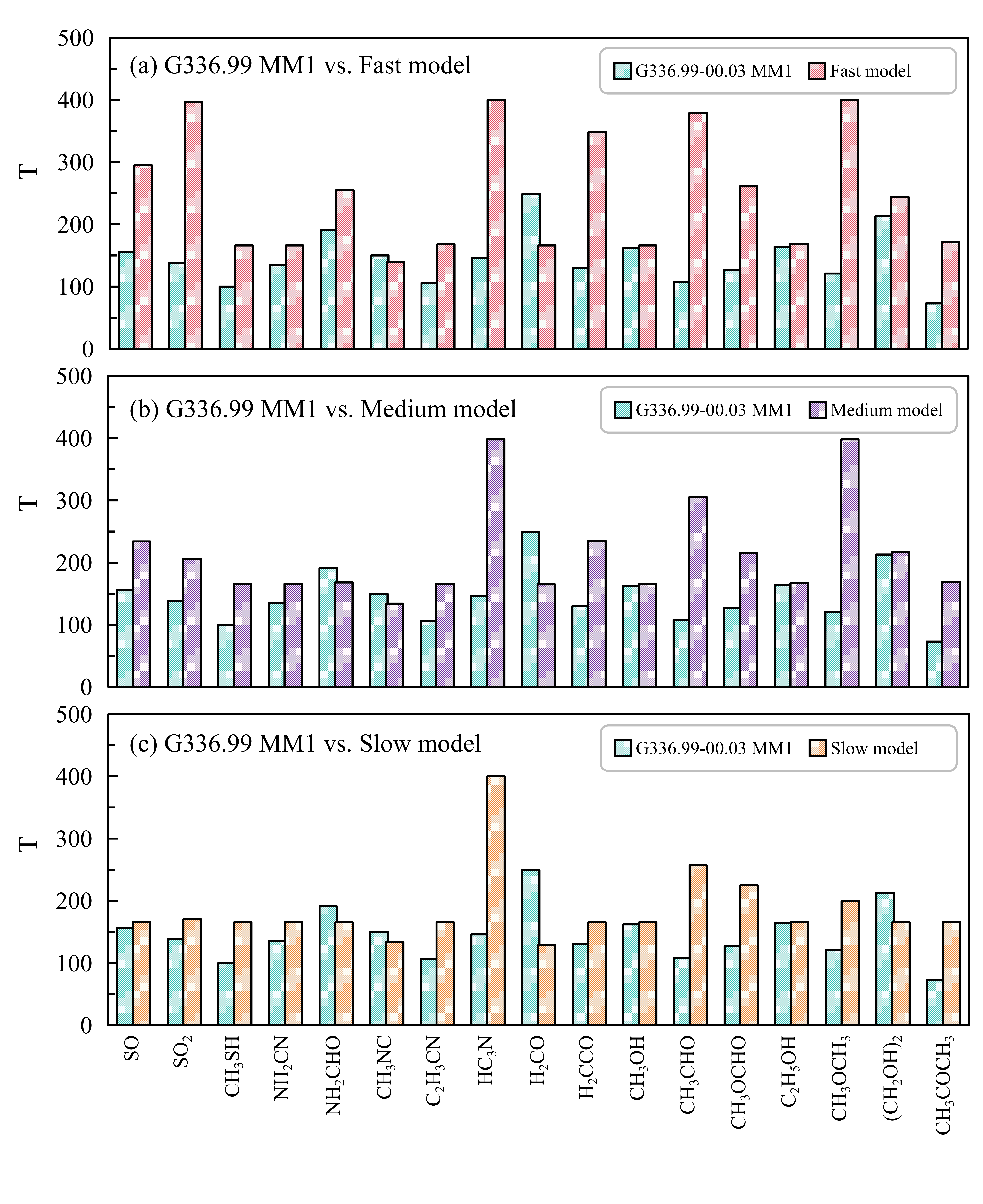}
\caption{Comparison of the observed molecular temperatures in G336.99-00.03 MM1 (cyan bars) with the temperatures corresponding to the peak gas-phase molecular abundances predicted by models from \citet{Gar22}. Colored bars represent the molecular temperatures derived from different warm-up timescale models.
\label{fig:fig. 15}}
\end{figure}


\bibliography{G336}{}
\bibliographystyle{aasjournal}



\end{document}